\documentclass[sigconf,authorversion]{acmart}
\acmSubmissionID{2317}

\usepackage[capitalize]{cleveref}
\usepackage{colortbl}
\usepackage{subcaption}
\usepackage{graphicx} 
\definecolor{best}{RGB}{255, 179, 179}
\definecolor{second}{RGB}{255, 219, 172}
\definecolor{third}{RGB}{255, 255, 204}

\crefname{section}{Sec.}{Secs.}
\Crefname{section}{Section}{Sections}
\crefname{table}{Tab.}{Tabs.}
\Crefname{table}{Table}{Tables}
\usepackage{enumitem}
\usepackage{balance}

\AtBeginDocument{%
  }

\copyrightyear{2025}
\acmYear{2025}
\setcopyright{acmlicensed}
\acmConference[SA Conference Papers '25]{SIGGRAPH Asia 2025 Conference Papers}{December 15--18, 2025}{Hong Kong, Hong Kong}
\acmBooktitle{SIGGRAPH Asia 2025 Conference Papers (SA Conference Papers '25), December 15--18, 2025, Hong Kong, Hong Kong}\acmDOI{10.1145/3757377.3763993}
\acmISBN{979-8-4007-2137-3/2025/12}

\acmJournal{TOG}

\begin{document}

\title{AD-GS: Alternating Densification for Sparse-Input 3D Gaussian Splatting}

\author{Gurutva Patle}
\affiliation{%
  \institution{Indian Institute of Science}
  \city{Bengaluru}
  \country{India}
}
\email{gurutvac@iisc.ac.in}

\author{Nilay Girgaonkar}
\affiliation{%
  \institution{Birla Institute of Technology and Science, Pilani}
  \city{Hyderabad}
  \country{India}
}
\email{f20210566@hyderabad.bits-pilani.ac.in}

\author{Nagabhushan Somraj}
\affiliation{%
  \institution{Indian Institute of Science}
  \city{Bengaluru}
  \country{India}
}
\email{nagabhushans@iisc.ac.in}

\author{Rajiv Soundararajan}
\affiliation{%
  \institution{Indian Institute of Science}
  \city{Bengaluru}
  \country{India}
}
\email{rajivs@iisc.ac.in}



\begin{teaserfigure}
  \centering
  \begin{subfigure}[t]{0.48\textwidth}
    \centering
    \includegraphics[width=\linewidth]{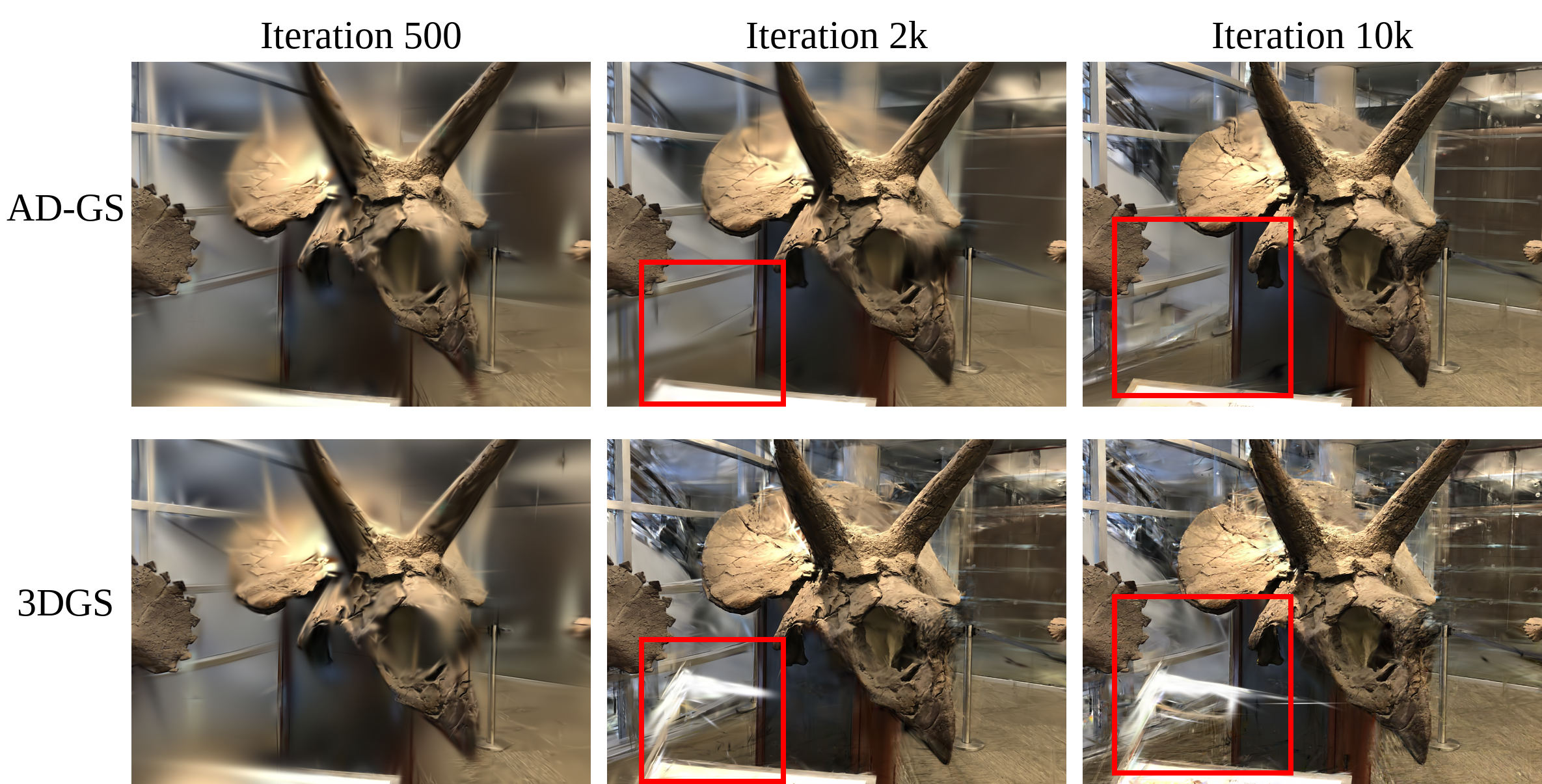}
    \caption{3DGS develops floaters (as seen in the red boxes) as we train for more iterations due to its uncontrolled densification. However, our AD-GS model is able to suppress floaters.}
    \label{fig:teaser_1}
  \end{subfigure}
  \hfill
  \begin{subfigure}[t]{0.49\textwidth}
    \centering
    \includegraphics[width=\linewidth]{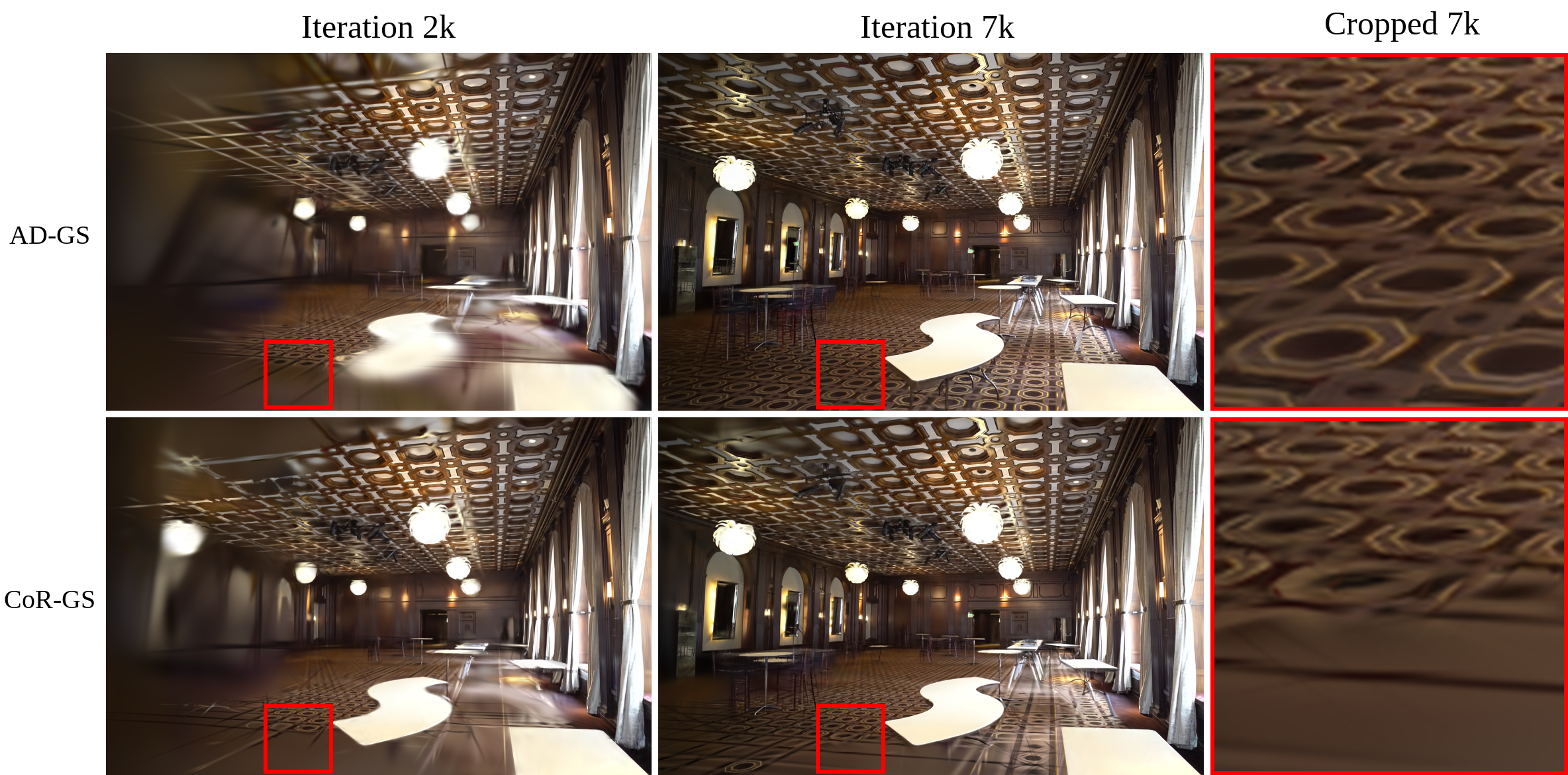}
    \caption{CoR-GS \cite{zhang2024cor} tends to employ aggressive smoothing to resolve floaters at the cost of loss of details, whereas AD-GS is able to reconstruct textures better.}
    \label{fig:teaser_2}
  \end{subfigure}
  \caption{AD-GS can simultaneously mitigate floater artifacts and retain high-frequency details in sparse input novel view synthesis.}
  \label{fig:teaser}
\end{teaserfigure}

\begin{abstract}
  3D Gaussian Splatting (3DGS) has shown impressive results in real-time novel view synthesis. However, it often struggles under sparse-view settings, producing undesirable artifacts such as floaters, inaccurate geometry, and overfitting due to limited observations. We find that a key contributing factor is uncontrolled densification, where adding Gaussian primitives rapidly without guidance can harm geometry and cause artifacts.
We propose \textbf{AD-GS}, a novel alternating densification framework that interleaves high and low densification phases. During high densification, the model densifies aggressively, followed by photometric loss based training to capture fine-grained scene details. Low densification then primarily involves aggressive opacity pruning of Gaussians followed by regularizing their geometry through pseudo-view consistency and edge-aware depth smoothness. This alternating approach helps reduce overfitting by carefully controlling model capacity growth while progressively refining the scene representation.
Extensive experiments on challenging datasets demonstrate that AD-GS significantly improves rendering quality and geometric consistency compared to existing methods. 
The source code for our model can be found on our project page:
 \url{https://gurutvapatle.github.io/publications/2025/ADGS.html}.
\end{abstract}



\maketitle

\section{Introduction}
Novel view synthesis is a fundamental problem in computer vision and graphics, with broad applications in 3D reconstruction and rendering. Neural Radiance Fields (NeRFs) \cite{mildenhall2020nerf} led to significant advances, but they suffer from slow rendering speeds. Recently, 3D Gaussian Splatting (3DGS) \cite{kerbl3Dgaussians} has emerged as an attractive alternative, enabling real-time rendering with high visual fidelity. 3DGS models a scene using a set of Gaussian primitives, where each Gaussian is parametrized by its 3D position, scale, color features, and opacity. These 3D Gaussians are then splatted on 2D and rendered to generate novel views. However, the performance of 3DGS rendering  substantially degrades when trained with sparse input views. Our work addresses the challenge of learning a 3DGS model given a few input views of a scene, where the limited number of input images makes geometry recovery and high-quality rendering particularly difficult. 

One of the major distortions that occur while training 3DGS models under sparse input conditions is the appearance of floating objects (or floaters) with incorrect depths. Several recent methods seek to improve 3DGS under sparse conditions by mitigating such distortions. FSGS \cite{zhu2024fsgs} incorporates priors from deep depth prediction networks. However, such depth priors may not generalize well for diverse scenes. CoR-GS \cite{zhang2024cor} enforces consistency between paired 3DGS models, but this tends to smooth out the reconstructions leading to loss of details. DropGaussian \cite{park2025dropgaussian} randomly drops Gaussians during training with increasing rates as training progresses, but this can hurt the reconstruction of fine-grained details. Thus, there is a need to address the shortcomings of existing sparse-input 3DGS methods in balancing the recovery of details and removing floaters. 

In sparse-view 3DGS, the limited number of input images severely limits the number of points in the initial point cloud leading to a poor initialization of the 3D Gaussians. A crucial step in 3DGS that allows the recovery of fine-details is the densification step, where new Gaussians are introduced through cloning or splitting. 
However, in sparse-view settings, uncontrolled densification may result in floaters due to the imprecise placement of newly introduced Gaussians. For example, when densification occurs due to splitting of existing Gaussians, the new Gaussians are sampled from the underlying distribution without constraints and can lead to poorly placed samples. 
These new wrongly placed Gaussians can lead to further cloning and densification around them. Moreover, since there are no explicit constraints on their positioning, the optimization does not remove these artifacts and the Gaussians may continue to remain in incorrect locations. Although the 3DGS densification step has attracted a lot of attention~\cite{rota2024revising,grubert2025improving}, and different losses have been explored to control densification, the primary goal has been to limit the model size. Its role in the sparse input case has hardly been explored. 

We propose to regulate the densification process by alternating between high and low densification. Specifically, during high densification, we use lower thresholds on the gradient magnitude of the loss at pixel locations, thereby enabling the introduction of a larger number of Gaussians in regions with high residual error. This phase aims to rapidly recover fine-grained scene details. This is followed by a period of recovery of scene details by training 3DGS with the photometric loss. There are no other constraints during this period of training and the focus is solely to enable the model retrieve fine-grained details. 

However, aggressive densification can introduce geometrically inconsistent Gaussians, potentially leading to artifacts such as floaters. To mitigate this, we perform opacity based pruning with elevated thresholds, eliminating low-opacity Gaussians that may cause artifacts. In addition, we introduce an explicit geometric regularization phase, wherein the model is trained not only with the photometric loss but also with additional geometric constraints. These include an edge-aware depth smoothness loss, which promotes local consistency in the underlying depth, and a pseudo-view consistency loss, applied between two jointly trained 3DGS models, to suppress geometric artifacts.

Our alternating densification framework is designed to enable controlled growth of the Gaussian set, allowing the model to progressively recover fine-grained scene details. At the same time, it ensures that newly introduced Gaussians are subject to geometric constraints, thereby reducing the likelihood of artifacts. Although one might consider applying geometric regularization throughout the entire training process, we find that doing so can inhibit the model’s ability to reconstruct intricate details, as the constraints may overly restrict the flexibility of the representation. In contrast, our intermittent application of geometry-guided regularization, strikes a more effective balance, enabling both the accurate recovery of visual detail and the mitigation of geometric artifacts as shown in \cref{fig:teaser}.

In summary, our main contributions are as follows:
\begin{itemize}
    \item An alternate densification framework involving high densification to enable recovery of fine-grained details and low densification involving aggressive pruning of Gaussians to mitigate artifacts.
    \item A geometry constrained training phase encouraging edge-aware depth smoothness and pseudo-view consistency to correct artifacts introduced during aggressive densification . 
    \item Significant improvements in novel view rendering under sparse input, with fewer floaters and richer reconstruction of scene details. 
\end{itemize}

\section{Related Work}


Novel view synthesis has been extensively studied in computer vision and graphics. Classical approaches include image based rendering assuming depth knowledge \cite{chen1993view} and light field rendering~\cite{levoy1996light}~\cite{gortler1996lumigraph}. With the advent of deep learning, learning-based approaches such as DeepStereo~\cite{flynn2016deepstereo} and multiplane image (MPI) based representations~\cite{zhou2018stereomag}  \cite{shih20203d} significantly improved the quality of novel renders.  The emergence of Neural Radiance Fields (NeRF)~\cite{mildenhall2020nerf} marked a turning point by modeling the entire 3D scene as a continuous function learned from multi-view images. 3DGS \cite{kerbl3Dgaussians} addresses the shortcomings of NeRFs with respect to their slow rendering and training times. Nevertheless, both NeRFs and 3DGS models degrade with sparse input when training on a given scene. In the following, we discuss prior work on sparse input NeRFs and 3DGS methods and then discuss how densification has been addressed in the 3DGS literature. 

\subsection{Sparse-Input NeRF}

To address sparse input constraints in NeRFs, an important line of work involves constraining the training with relevant priors. Depth-based priors are shown to be effective in supervising NeRFs under limited views. RegNeRF~\cite{niemeyer2022regnerf} and DSNeRF~\cite{deng2022dsnerf} introduce supervision through depth smoothness or sparse COLMAP-based depth estimates. DDP-NeRF uses deep network based depth estimators as priors. While ViP-NeRF \cite{somraj2023vipnerf} employs priors on relative depth,  SimpleNeRF~\cite{somraj2023simplenerf} introduces auxiliary models with lower modeling capacity to guide the main NeRF by providing reliable depth supervision in selected locations. In contrast to depth based regularization, FreeNeRF~\cite{yang2023freenerf} explores frequency regularization by gradually increasing the ability of the model to capture high-frequency details as training progresses. 
On the other hand, \cite{chen2022hallucinated} hallucinates and imposes semantic consistency with the main model for regularization. Despite rich literature in sparse-input NeRFs, all these methods suffer from the inherent limitations of NeRFs with respect slow training and rendering speeds. Thus, there is a need to study sparse-input 3DGS to allow for fast training and rendering models. Further, many priors used for sparse input NeRF are not directly applicable for 3DGS. There is a need to design priors specifically for 3DGS models.   

\subsection{Sparse-Input 3D Gaussian Splatting}

Recent extensions of 3DGS to sparse-input regimes regularize the reconstruction through various priors. FSGS~\cite{zhu2024fsgs} uses depth maps predicted by pretrained networks \cite{Ranftl2021}  \cite{Ranftl2022}, while CoR-GS~\cite{zhang2024cor} leverages pseudo-view consistency to minimize disagreement between two independently trained Gaussian fields. Coherent-GS~\cite{paliwal2024coherentgs} encourages Gaussians to optimize and position themselves as coherent entities by providing segmentation masks as input. 
On the other hand, InstantSplat \cite{fan2024instantsplat} employs stereo-depth estimates from pre-trained networks \cite{leroy2024grounding} for rich initalization. \cite{jiang2024construct} instead uses monocular depth estimates for initialization. Both these models allow for pose-free sparse input 3DGS. 
In contrast to explicit priors, DropGaussian \cite{park2025dropgaussian} reduces overfitting to the input views by randomly dropping Gaussians and employing edge-guided splitting.  
Different from the above methods, we study how densification has a significant impact on the performance of 3DGS in the sparse input case. 

Feed-forward 3DGS methods represent an alternative direction, where neural networks are trained on a corpus of scenes. PixelSplat~\cite{charatan2024pixelsplat} and MVSplat~\cite{chen2024mvsplat} learn deep networks to predict Gaussian parameters from few input views. Recent approaches such as  FLARE~\cite{zhang2025flare} and NoPoseSplat~\cite{ye2024no} also allow pose-free training. However, these methods are typically limited by low-resolution outputs and such methods may not generalize to diverse scene types. 

\subsection{Controlling Densification in 3DGS}

Densification plays a crucial role in 3DGS, in converting a sparse point cloud to detailed pixel reconstructions. Several prior works focus on controlling densification in the dense input case primarily to reduce memory and computational overhead. In particular, ~\citet{rota2024revising}, ~\citet{zhang2024pixel}, and ~\citet{grubert2025improving} explore criteria such as local image cues, occupancy maps, and projection overlap to densify the scene. HDA-GS~\cite{wang2025hda} introduces hierarchical density adaptation by splitting Gaussians in sparsely populated point cloud regions to Gaussians with low volume and high opacity.
However, these improvements are not targeted at sparse-input settings and do not explicitly address artifacts such as floaters or overfitting. In contrast, our work demonstrates that controlling densification can simultaneously improve details and suppress artifacts in the sparse-input regime.

\begin{figure}[ht]
    \centering
    \includegraphics[width=\linewidth]{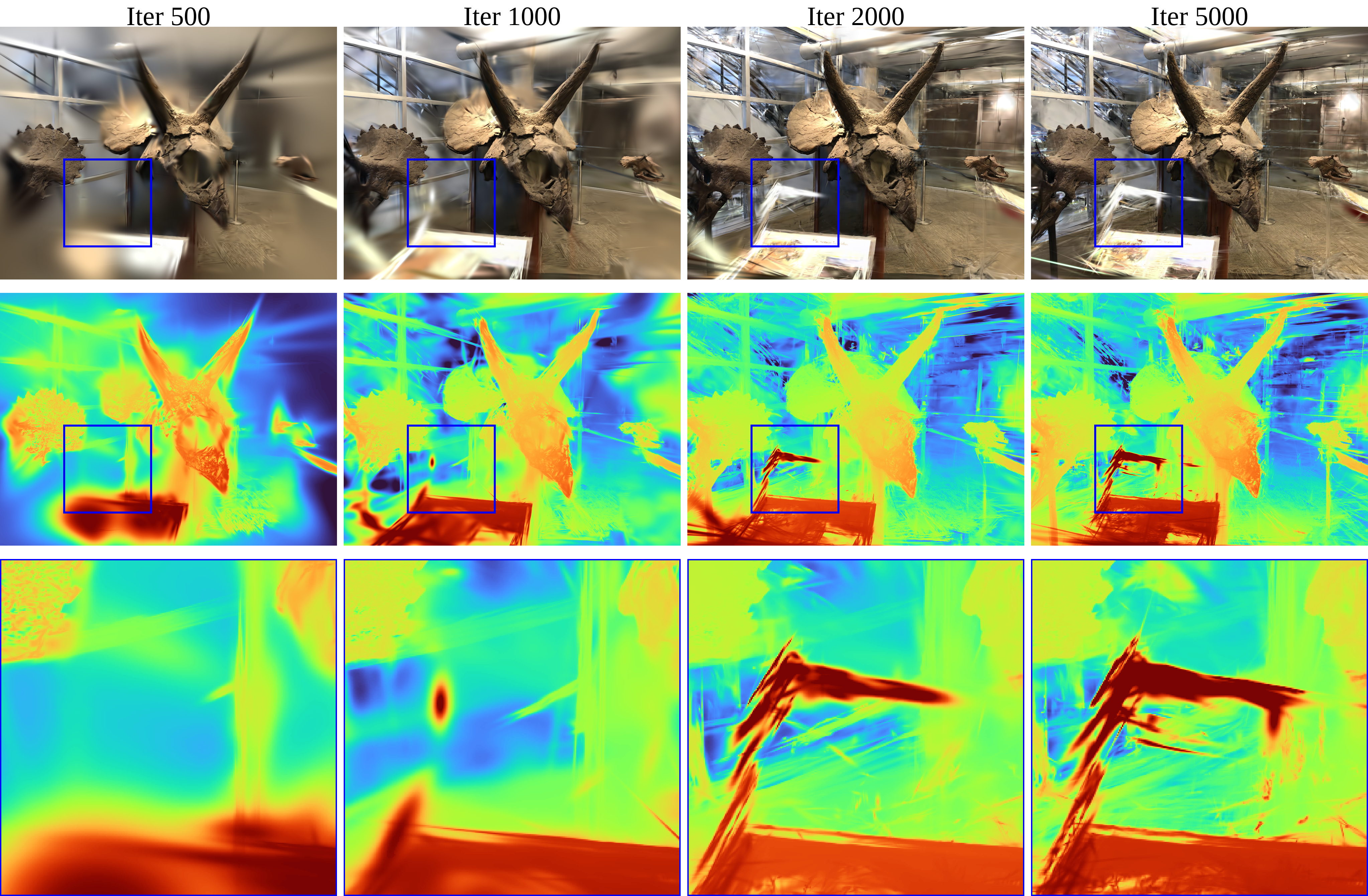}
    \caption{As training progresses, densification under sparse views leads to floating artifacts and inconsistent geometry.}
    \label{fig:densification-failure}
\end{figure}

\begin{figure*}
        \centering
        \includegraphics[width=0.995\linewidth]{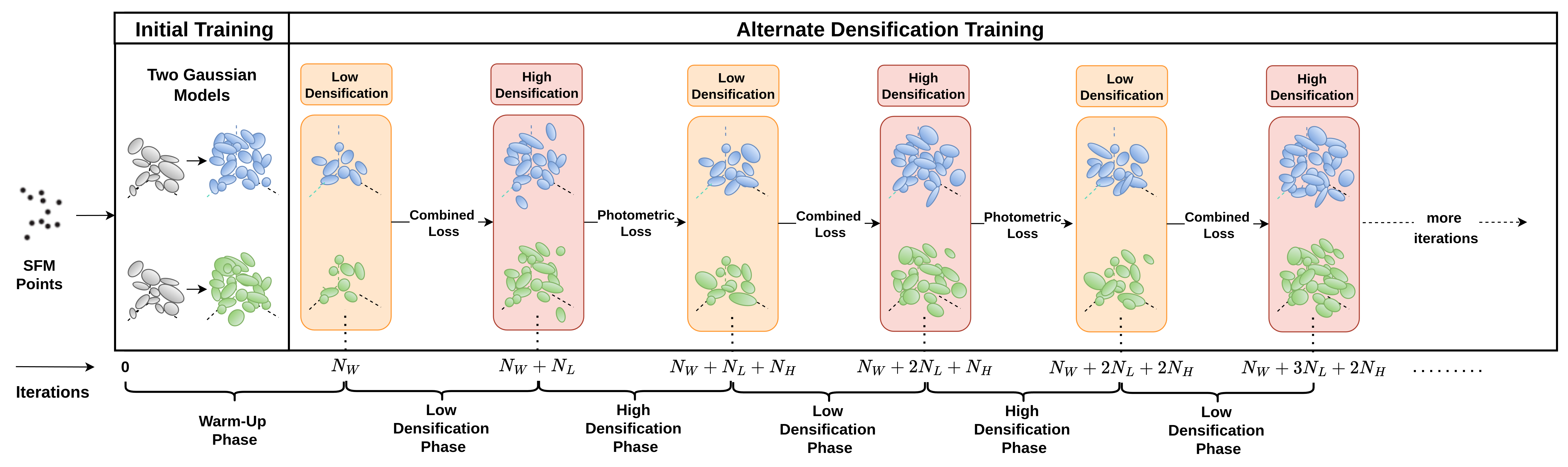}
        \caption{\textbf{AD-GS Method}: The training begins with a Warm-Up Phase of $N_W$ iterations, where two 3DGS models 
        are trained independently using photometric loss. This is followed by the Alternating Densification Phase, which alternates between low and high densification steps every $N_L$ and $N_H$ iterations. Low Densification applies strict gradient and opacity thresholds and includes additional supervision via pseudo-view consistency and edge-aware depth smoothness losses (we define the total loss in Equation (\ref{eq:ldtotalloss}) during this phase as "Combined Loss"). High Densification enables aggressive Gaussian growth and uses only photometric loss 
        to recover high-frequency details. Over iterations, this alternation progressively refines scene geometry while avoiding overfitting and floaters. The figure shows how Gaussian counts increase in high densification and decrease due to pruning in slow densification.
        }
        \label{fig_qualitative-realestate02}
\end{figure*}

\section{Method}\label{sec:method}

\subsection{Preliminaries: 3D Gaussian Splatting}\label{method:3dgs}

3D Gaussian Splatting represents a scene using a set of 3D Gaussians. Each Gaussian is parameterized by a center position $\boldsymbol{\mu} \in \mathbb{R}^3$, a scale vector $\mathbf{s} \in \mathbb{R}^3$, a rotation quaternion $\mathbf{q} \in \mathbb{R}^4$, an opacity scalar $o \in [0, 1]$, and a color feature vector $\mathbf{f} \in \mathbb{R}^K$. The color feature $\mathbf{f}$ is typically represented using spherical harmonics (SH). The scale and rotation define a 3D covariance matrix $\boldsymbol{\Sigma}$ as:
\begin{equation}
    \boldsymbol{\Sigma} = R S S^\top R^\top,
\end{equation}
where $R$ is the rotation matrix derived from the quaternion $\mathbf{q}$, and $S = \mathrm{diag}(s_x, s_y, s_z)$ is the scaling matrix. Each Gaussian defines a spatial density at position $x \in \mathbb{R}^3$ around its center $\mu$ as:
\begin{equation}
    G(x) = \exp\left(-\frac{1}{2}(x - \mu)^\top \Sigma^{-1} (x - \mu)\right),
    \label{eq:gaussian-density}
\end{equation}
To render Gaussians into an image, we project each 3D Gaussian onto the image plane as a 2D Gaussian under a camera transformation. This transformation is characterized by a camera matrix $W$, and the 2D projected covariance $\Sigma^{2D}$ is given by:
\begin{equation}
    \Sigma^{2D} = J W \Sigma W^\top J^\top,
    \label{eq:covariance2d}
\end{equation}
where $J$ is the Jacobian of the local affine approximation of the projection.
To render an image from a camera view, each 3D Gaussian is projected to the image plane. For pixel rendering, 3DGS sorts the 3D Gaussian primitives intersecting a pixel’s camera ray. The final color of a pixel $p$ is obtained via alpha compositing of the contributions from the projected 2D Gaussians overlapping that pixel as 
\begin{equation}
    C(p) = \sum_{i=1}^{N(p)} c_i \cdot \alpha_i \cdot \prod_{j=1}^{i-1} (1 - \alpha_j).
    \label{eq:rendered-color}
\end{equation}
where $N(p)$ is the number of Gaussians visible at pixel $p$. Here, $c_i$ and $\alpha_i$ represent the effective color and opacity of the $i$-th Gaussian. The color $c_i$ is computed from the SH coefficients based on the view direction, while $\alpha_i$ is derived from the learnable opacity, shape and position of the projected Gaussian.
3DGS also employs densification (through splitting and cloning) in addition to pruning to adaptively regulate the number and density of
Gaussian primitives.

\subsection{Analysis of Densification under Sparse Input}\label{subsec:densification-problems}

In 3D Gaussian Splatting, Gaussians are densified based on the 3D gradients of the photometric loss. Typically, densification is triggered every fixed number of iterations, where Gaussians with error gradients greater than a threshold $\tau_{\text{pos}}$, are either split or cloned depending on their respective scales.
Gaussians are cloned in under-reconstructed regions to add more detail by creating another copy of the same Gaussian. On the other hand, large Gaussians are split by creating two Gaussians sampled from the parent Gaussian distribution with reduced scale parameters. 
While this process is effective in capturing scene details, it is inherently stochastic and uncontrolled. This randomness especially during splitting, can introduce Gaussians in geometrically inconsistent or low-likelihood regions, particularly in sparse-view settings. 

As training progresses, such Gaussians may start overfitting to the limited input views, attempting to match observed pixels even when geometrically incorrect. This results in the emergence of floating artifacts, or \textit{floaters}, which visually detach from the scene structure and degrade rendering quality.
Vanilla 3DGS, without explicit supervision or regularization, fails to eliminate these erroneous Gaussians. The pruning operation, which removes Gaussians based on low opacity or excessive size (for example $\alpha_i < \epsilon_\alpha$ for the $i^{\text{th}}$ Gaussian), is often too ineffective in sparse settings.
The floaters once formed are not removed by the photometric loss. Instead, the color of the floaters is changed to match the sparse input views, leading to distortions in novel views. \cref{fig:densification-failure} illustrates this phenomenon by showing how floaters gradually emerge over training in vanilla 3DGS, highlighting the need for a more controlled and geometry-aware densification strategy.

\begin{table*}
\centering
\caption{Quantitative results on the Tanks \& Temples dataset under different view settings. $\uparrow$ indicates higher is better, $\downarrow$ indicates lower is better. \colorbox{best}{Best}, \colorbox{second}{Second Best}, and \colorbox{third}{Third Best} results are highlighted.}
\label{tab:tanks_metrics}

\vspace{1mm}
\textbf{Tanks \& Temples } \\
\begin{tabular}{l|ccc|ccc|ccc}
\toprule
\textbf{Model} 
& \multicolumn{3}{c|}{PSNR $\uparrow$} 
& \multicolumn{3}{c|}{SSIM $\uparrow$} 
& \multicolumn{3}{c}{LPIPS $\downarrow$} \\
& 3-view & 6-view & 9-view 
& 3-view & 6-view & 9-view 
& 3-view & 6-view & 9-view \\
\midrule
3DGS         & 16.992 & 22.393 & 24.492 & 0.556 & 0.788 & 0.850 & 0.352 & 0.182 & 0.141 \\
FSGS         & \cellcolor{second}19.644 & \cellcolor{third}26.436 & \cellcolor{third}28.488 & 0.637 & 0.847 & \cellcolor{third}0.888 & \cellcolor{second}0.312 & \cellcolor{second}0.168 & 0.139 \\
CoR-GS       & 19.246 & 26.273 & \cellcolor{second}28.489 & \cellcolor{third}0.650 & \cellcolor{third}0.849 & \cellcolor{second}0.891 & 0.342 & 0.175 & \cellcolor{second}0.132 \\
DropGaussian & \cellcolor{third}19.475 & \cellcolor{second}26.452 & \cellcolor{third}28.488 & \cellcolor{second}0.652 & \cellcolor{second}0.852 & \cellcolor{second}0.891 & \cellcolor{third}0.320 & \cellcolor{third}0.173 & \cellcolor{third}0.138 \\
\textbf{AD-GS (Ours)} & \cellcolor{best}19.724 & \cellcolor{best}27.139 & \cellcolor{best}28.964 & \cellcolor{best}0.678 & \cellcolor{best}0.867 & \cellcolor{best}0.902 & \cellcolor{best}0.285 & \cellcolor{best}0.142 & \cellcolor{best}0.113 \\
\bottomrule
\end{tabular}
\end{table*}

\begin{table*}
\centering
\caption{Quantitative results on the LLFF dataset under different view settings.}
\label{tab:llff_metrics}

\textbf{LLFF} \\
\begin{tabular}{l|ccc|ccc|ccc}
\toprule
\textbf{Model} 
& \multicolumn{3}{c|}{PSNR $\uparrow$} 
& \multicolumn{3}{c|}{SSIM $\uparrow$} 
& \multicolumn{3}{c}{LPIPS $\downarrow$} \\
& 3-view & 6-view & 9-view 
& 3-view & 6-view & 9-view 
& 3-view & 6-view & 9-view \\
\midrule
3DGS         
& 18.90 & 22.57 & 24.02 
& 0.632 & 0.757 & 0.799 
& \cellcolor{third}0.269 & \cellcolor{second}0.182 & \cellcolor{third}0.151 \\

FSGS         
& 19.50 & 23.25 & 24.56 
& \cellcolor{third}0.655 & 0.776 & 0.815 
& \cellcolor{third}0.269 & \cellcolor{third}0.185 & \cellcolor{third}0.151 \\

CoR-GS       
& \cellcolor{third}19.59 & \cellcolor{third}23.33 & \cellcolor{third}24.75 
& \cellcolor{second}0.674 & \cellcolor{third}0.780 & \cellcolor{third}0.817 
& 0.271 & 0.187 & 0.152 \\

DropGaussian 
& \cellcolor{second}19.72 & \cellcolor{second}23.52 & \cellcolor{second}24.89 
& \cellcolor{second}0.674 & \cellcolor{second}0.786 & \cellcolor{second}0.824 
& \cellcolor{second}0.266 & \cellcolor{second}0.182 & \cellcolor{second}0.148 \\

\textbf{AD-GS (Ours)} 
& \cellcolor{best}20.06 & \cellcolor{best}23.54 & \cellcolor{best}25.02 
& \cellcolor{best}0.699 & \cellcolor{best}0.793 & \cellcolor{best}0.830 
& \cellcolor{best}0.237 & \cellcolor{best}0.170 & \cellcolor{best}0.142 \\
\bottomrule
\end{tabular}
\end{table*}

\subsection{Alternate Low and High Densification}





To address the above challenges, we introduce \textbf{AD-GS}, a training framework for Gaussian Splatting that progressively enhances scene quality by alternating between low and high densification phases. 
This alternating strategy keeps model growth under control while incrementally enhancing geometry and appearance, guided by both photometric and geometric constraints, under sparse supervision. The training process consists of three primary phases: an initial warm-up phase, followed by alternating low and high densification phases.
To enable pseudo-view consistency as a form of regularization during the low densification phase, we train two separate 3DGS models concurrently \cite{zhang2024cor}. These two models are initialized independently and optimized simultaneously throughout training.
For evaluation, we use one of the two trained 3DGS models (chosen arbitrarily) as our final test-time model.

\subsubsection{Warm-Up Phase (Initial Training)} 
The warm-up phase spans the first $N_W$ iterations and is designed to build a stable initialization. Both the 3DGS models are trained independently using the same loss similar to 3DGS method.  In particular, let the train views from a given viewpoint rendered by 3DGS models $G_1$ and $G_2$ be $v_1$ and $v_2$ respectively. The photometric loss for $G_k, k\in\{1,2\}$ is defined as
\begin{equation}\label{eq:photometric}
L_{\text{ph}}(v_k, v_{gt}) = (1 - \lambda_{\text{ssim}}) \lVert v_k - v_{gt} \rVert_1 + \lambda_{\text{ssim}} (1 - \text{SSIM}(v_k, v_{gt})),
\end{equation}

\noindent where \(v_{gt}\) is the ground-truth view and \( \lambda_{\text{ssim}}\in\left[0,1\right]\) defines the weight for the two losses.

\begin{figure*}
        \centering
        \includegraphics[width=0.995\linewidth]{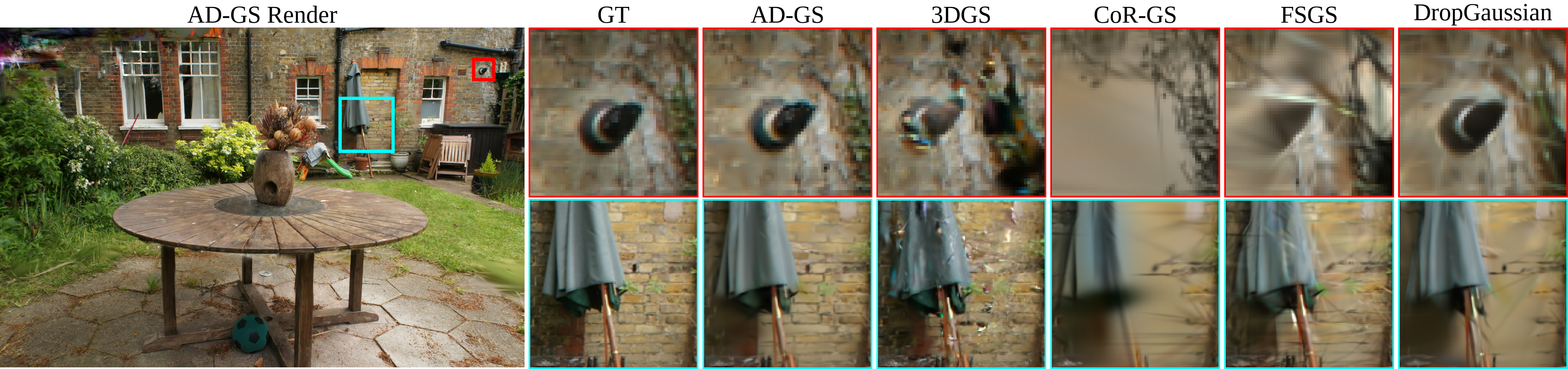}
        \caption{\textbf{Qualitative examples on the Mip-NeRF360 dataset (Garden scene)} with 24 input views. AD-GS reconstructs the videocamera and the texture in the background better than the other models.
        }
        \label{fig:results-m360-24}
    \end{figure*}
    
    \begin{figure*}[t]
    \centering
    \includegraphics[width=0.995\linewidth]{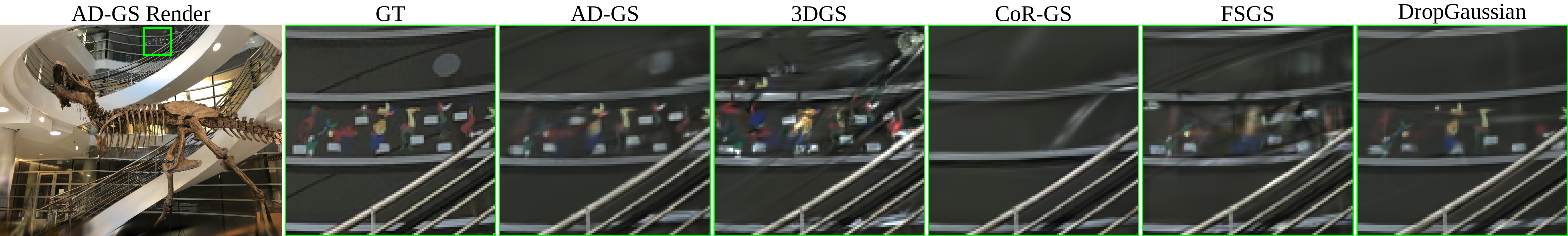}
    \caption{\textbf{Qualitative example on the LLFF dataset (T-Rex scene)} with 3 input views. AD-GS manages to reconstruct fine-grained background details, while other models fail to do so. In particular, the 3DGS reconstruction is noisy, while CoR-GS suffers from excessive smoothing.}
    \label{fig:results-llff-3}
\end{figure*}

\subsubsection{Alternating Densification Phases}
After warm-up, we alternate between high and low densification phases designed to achieve different objectives.

\vspace{1em}

\textbf{Low Densification Phase}
The low densification phase consists of a period of $N_L$ iterations where there is densification in one iteration followed by geometry-constrained training in all iterations of this phase. 
In particular, the low densification step consists of two key steps. Firstly, we use an elevated opacity threshold $\epsilon_L>\epsilon_{\alpha}$ to aggressively remove low opacity Gaussians and limit overfitting. This helps remove translucent floaters. Secondly, we use a strict gradient norm threshold $\tau_L>\tau_{\text{pos}}$ to densify at a slower rate. These two strategies help limit the number of Gaussians that clone or split in this step. 

To enforce geometric consistency and regularization, we enable pseudo-view consistency and edge-aware depth smoothness constraints. These constraints encourage Gaussians to align with plausible surfaces and discourage isolated or floating Gaussians.

\paragraph{(a) Pseudo-View Consistency:}
We render a pseudo-view that is not part of the training views from the two 3DGS models to get $u_1$ and $u_2$. The pseudo-view is chosen by adding a small perturbation to one of the training views. Pseudo-view consistency involves photometric consistency between the rendered views of the two models. In particular, the loss is defined as
\begin{equation}
L_{\text{pseudo}}(u_1,u_2) = L_{\text{ph}}(u_1, u_2),
\end{equation}
where $L_{\text{ph}}$ is defined in Equation (\ref{eq:photometric}).
This loss ensures that both models converge towards consistent representations. Although this loss has been used in CoR-GS \cite{zhang2024cor} to constrain the models, we believe that such consistency regularization encourages smooth output similar to how consistency of perturbed observations leads to smoothed output in semi-supervised learning \cite{miyato2018virtual}. 
Thus, we employ such a loss to enable geometry consistency and mitigate floater artifacts. Note that employing such a loss throughout the training can inhibit the learning of fine-grained details as shown in \cref{fig:teaser_2} and Table \ref{tab:ablation_study_ssim_compact}. However, AD-GS applies ${L_{\text{pseudo}}}$ only during the low-densification phase, jointly with edge-aware depth smoothness.

\paragraph{(b) Edge-Aware Depth Smoothness:}
Since floaters can often be detected as instances where depth smoothness is violated, we encourage the model to achieve smooth depth maps while preserving discontinuities at intensity edges. To obtain depth smoothness, we first estimate the depth obtained at each pixel in a given view by the 3DGS model through rendering. If $z_i$ is depth of a $i^{\text{th}}$ Gaussian contributing to a pixel, then the depth of a pixel $p$ is obtained via the same alpha compositing as in Equation ~\eqref{eq:rendered-color} by replacing color $c_i$ with $z_i$.
For each 3DGS model, along with the rendered view $v$ for a given train viewpoint and pseudo-view $u$, we also get the corresponding rendered depths $d$ and $d_{u}$.

We define the edge-aware depth-smoothness loss in view $v$ as
\begin{equation}
L_{\text{ds}}(d, v) = 
\sum_{x, y} \lVert \nabla d(x, y) \rVert_1 
 \exp\left(- \lVert \nabla v(x, y) \rVert_1\right) 
- \lambda_{\text{r}} \left( d_{\text{max}} - d_{\text{min}} \right),
\end{equation}
where  $d(x, y)$ is the rendered depth,
$v(x, y)$ is the image intensity,
$\nabla d(x, y)$ is the spatial depth gradient, and
$\nabla v(x, y)$ is the spatial image gradient summed over all the color channels at pixel $(x, y)$.
Further, $\lambda_{\text{r}}>0$ is the regularization weight, while
$d_{\text{max}}$ and  $d_{\text{min}}$ are the maximum and minimum depth values in the rendered view. While the first term in the above loss encourages smoothness, the second term ensures that the depth values have a good dynamic range. 
The above loss penalizes depth changes more in smooth regions and less around edges, preventing over-smoothing. We impose this loss in both the train views as well as the pseudo-views. Thus, the total edge-aware depth-smoothness loss for training $G_k$, $k=1,2$ is:
\begin{equation}
L_{\text{tds}}(v_k, d_k,u_k, d_{\text{uk}}) = \omega_1 L_{\text{ds}}(v_k, d_k) + \omega_2 L_{\text{ds}}(u_k,d_{uk} ),
\end{equation}
where $\omega_1$ and $\omega_2$ are weights balancing both the smoothness terms. Note that one train view and pseudo-view are sampled in the above loss. Instead of relying on external depth priors like in \cite{zhu2024fsgs,paliwal2024coherentgs,fan2024instantsplat,jiang2024construct}, we regularize the rendered depth map by promoting smoothness in textureless regions while allowing for discontinuities at image intensity edges. 

The overall loss for training $G_k$, $k=1,2$, during the low densification phase is:
\begin{equation}\label{eq:ldtotalloss}
L_k = \lambda_1  L_{\text{ph}}(v_k,v_{gt}) + \lambda_2  L_{\text{tds}}(v_k, d_k,u_k, d_{uk})\ + \lambda_3  L_{\text{pseudo}}(u_1,u_2),
\end{equation}
where $\lambda_1$, $\lambda_2$ and $\lambda_3$ are used to weight the different losses. 

\vspace{1em}

\textbf{High Densification Phase}
To enable the model to learn high-frequency textures and fine-grained details, we switch to a high densification phase consisting of $N_H$ iterations, which again consists of one densification step along with vanilla 3DGS model parameter updates in all iterations of this phase. 
 In particular, we use an aggressive densification strategy, where the norm threshold for Gaussians to split or clone is set at $\tau_{pos}$. We apply such a densification for both models. We employ only the photometric loss and remove the regularization terms in this phase. Thus, the model learns to better fit the photometric cues from the training views and recover fine details.

     \begin{figure*}
        \centering
        \begin{minipage}{\linewidth}
        \centering
        \includegraphics[width=0.995\linewidth]{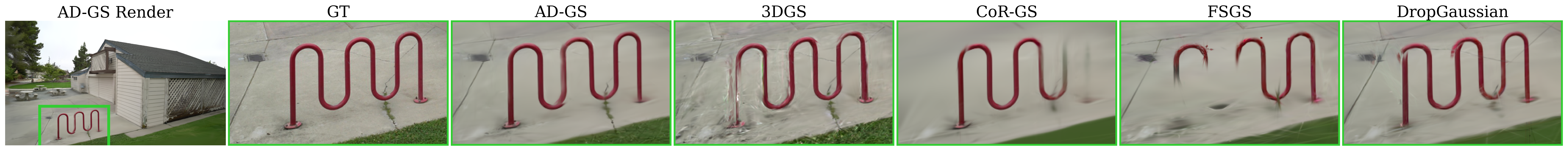}
    \end{minipage}

    \begin{minipage}{\linewidth}
        \centering
        \includegraphics[width=0.995\linewidth]{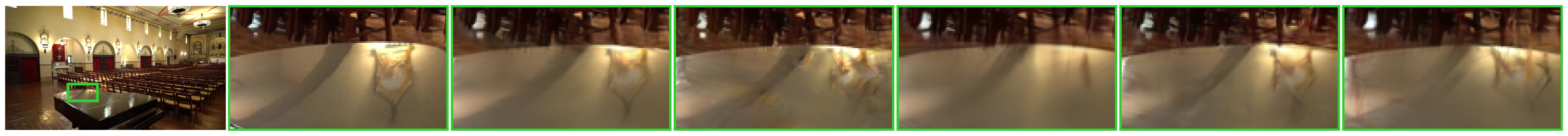}
    \end{minipage}
        \caption{\textbf{ Qualitative examples on the Tanks \& Temples dataset (Barn (top) and Church (bottom) scene)} with 3 input views. AD-GS reconstructs the structure with finer details and surfaces better than the other models.
        }
        \label{fig:tanks-3-barn-churc}
    \end{figure*}

    \begin{figure*}
    \centering
    \includegraphics[width=0.995\linewidth]{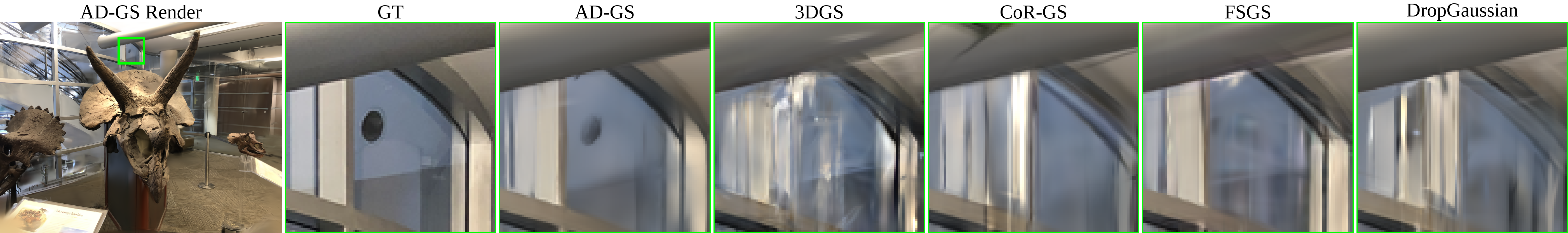}
    \caption{\textbf{Qualitative example on the LLFF dataset (Horns scene)} with 3 input views. The AD-GS better reconstructs the window and the details of the fixture behind it.}
    \label{fig:llff-3-horns}
\end{figure*}

    \begin{figure*}
        \centering
        \includegraphics[width=0.995\linewidth]{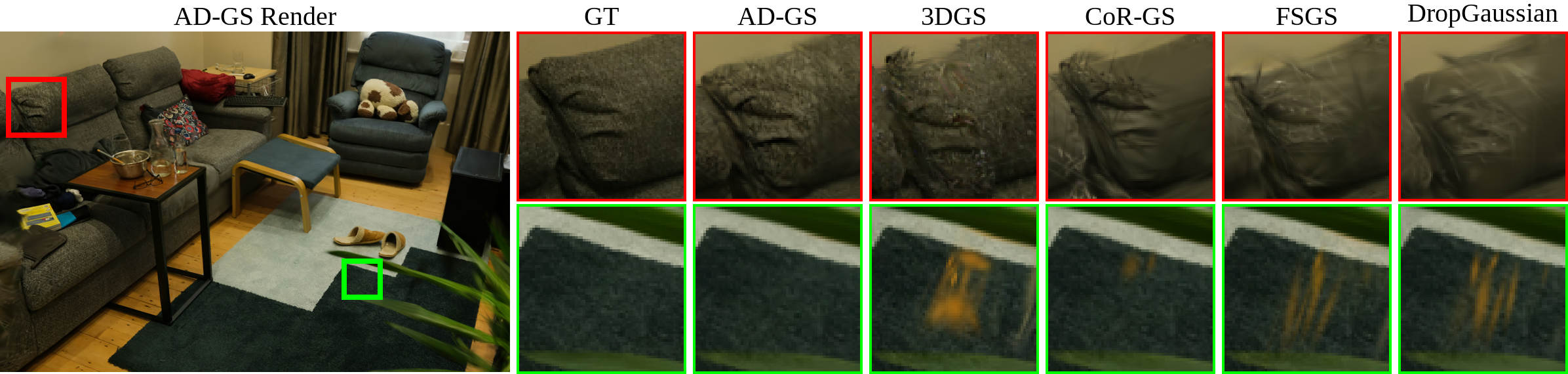}
        \caption{\textbf{Qualitative examples on the Mip-NeRF360 dataset (Room scene)} with 24 input views. AD-GS reconstructs the pillow and carpet texture, better than the competing models. The orange colored artifacts in other models can be seen as floater artifacts in other models.  
        }
        \label{fig:m360-24-room}
    \end{figure*}

\begin{table}[ht] 
\small
\centering
\caption{Quantitative results on the Mip-NeRF360 dataset under different view settings.}
\label{tab:m360-metrics}

\footnotesize
\textbf{Mip-NeRF360} \\
\begin{tabular}{l|cc|cc|cc}
\toprule
\textbf{Model} 
& \multicolumn{2}{c|}{PSNR $\uparrow$} 
& \multicolumn{2}{c|}{SSIM $\uparrow$} 
& \multicolumn{2}{c}{LPIPS $\downarrow$} \\
& 12-view & 24-view 
& 12-view & 24-view 
& 12-view & 24-view \\
\midrule
3DGS             
& 17.36 & 22.06 
& 0.496 & 0.701 
& \cellcolor{second}0.403 & \cellcolor{second}0.253 \\

FSGS             
& 18.51 & 23.11 
& 0.547 & 0.720 
& 0.411 & 0.275 \\

CoR-GS           
& \cellcolor{second}19.42 & \cellcolor{third}23.20 
& \cellcolor{second}0.579 & \cellcolor{third}0.728 
& \cellcolor{third}0.410 & \cellcolor{third}0.272 \\

DropGaussian     
& \cellcolor{third}19.18 & \cellcolor{second}23.28 
& \cellcolor{third}0.575 & \cellcolor{second}0.732 
& 0.412 & 0.277 \\

\textbf{AD-GS(Ours)}         
& \cellcolor{best}19.66 & \cellcolor{best}23.68 
& \cellcolor{best}0.593 & \cellcolor{best}0.750 
& \cellcolor{best}0.386 & \cellcolor{best}0.249 \\
\bottomrule
\end{tabular}
\end{table}

\section{Experiments}

\subsection{Experimental Settings}
We conduct experiments on the  LLFF~\cite{mildenhall2019llff}, Mip-NeRF360~\cite{barron2022mipnerf360}, and Tanks \& Temples~\cite{knapitsch2017tanks} datasets. Our train-test data splits are the same as that in previous studies \cite{zhang2024cor, zhu2024fsgs, park2025dropgaussian}. 
For LLFF and Tanks \& Temples, we train with 3, 6, and 9 input views per scene. For Mip-NeRF360, we report results with 12 and 24 training views. We assume known camera intrinsics and extrinsics, which is relevant in applications such as remote presence, and virtual exploration, where other sensors may be deployed to obtain camera parameters.

To assess model performance, we compute structural similarity index (SSIM), peak signal to noise ratio (PSNR), and learned perceptual image patch similarity (LPIPS) between the rendered images and the ground-truth target views.

\begin{figure*}
        \centering
        \includegraphics[width=0.995\linewidth]{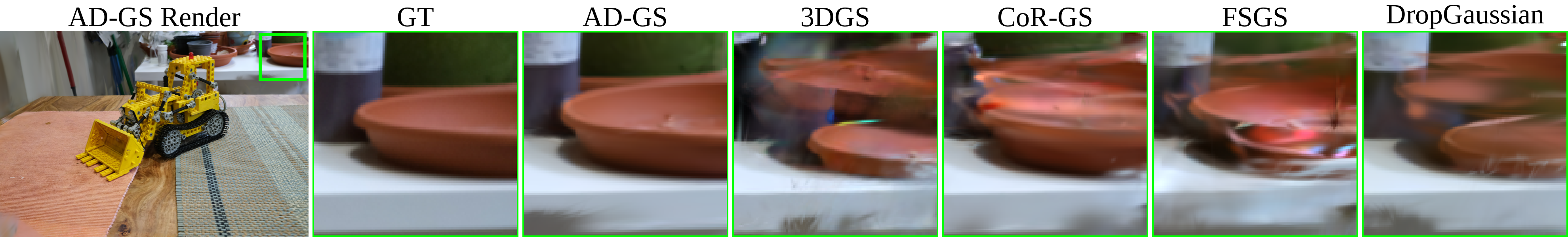}
        \caption{\textbf{Qualitative example on the Mip-NeRF360 dataset (Kitchen scene)} with 12 input views. The AD-GS prediction is closest to the ground truth; the other models suffer from ghosting artifacts.
        }
        \label{fig:results-m360-12}
    \end{figure*}

 \begin{figure*}
        \centering
        \includegraphics[width=0.995\linewidth]{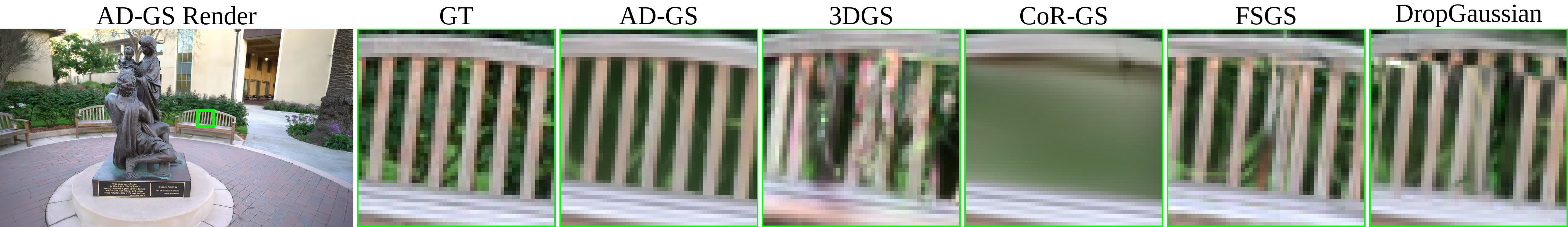}
        \caption{\textbf{Qualitative examples on the Tanks \& Temples dataset (Family scene)} with 3 input views. AD-GS replicates the geometry of the bench accurately.
        }
        \label{fig:tanks-3-family}
    \end{figure*}

\begin{figure*}
        \centering
        \includegraphics[width=0.995\linewidth]{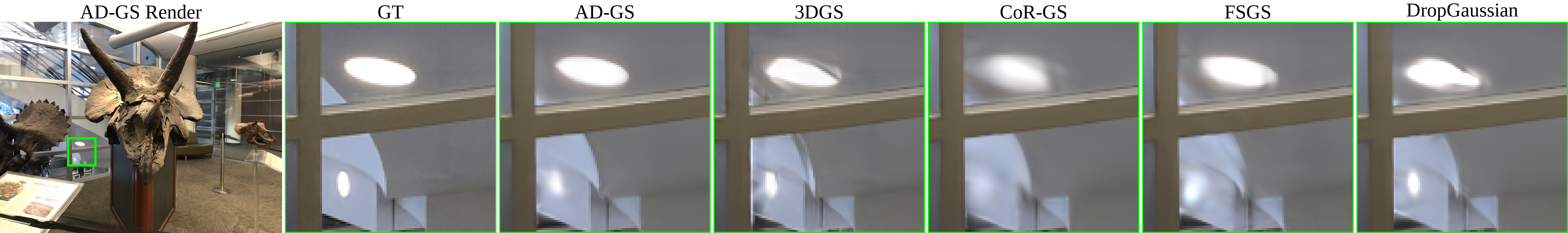}
        \caption{\textbf{Qualitative examples on the LLFF dataset (Horns Scene)} with 6 input views. AD-GS reconstructs the shape of the light fixture better than the other models.
        }
        \label{fig:llff-6-horns}
    \end{figure*}

\subsection{Comparisons}
We evaluate the performance of our proposed method against various sparse input 3DGS models such as FSGS \cite{zhu2024fsgs}, CoR-GS \cite{zhang2024cor}, and DropGaussian \cite{park2025dropgaussian}, along with vanilla 3DGS on all three datasets. All models are trained on datasets using the code provided by the respective authors. Implementation details and hyperparameters are provided in the supplementary. Importantly, all hyperparameters are fixed across all datasets and scenes.

\begin{figure}
    \centering
    \includegraphics[width=0.95\linewidth]{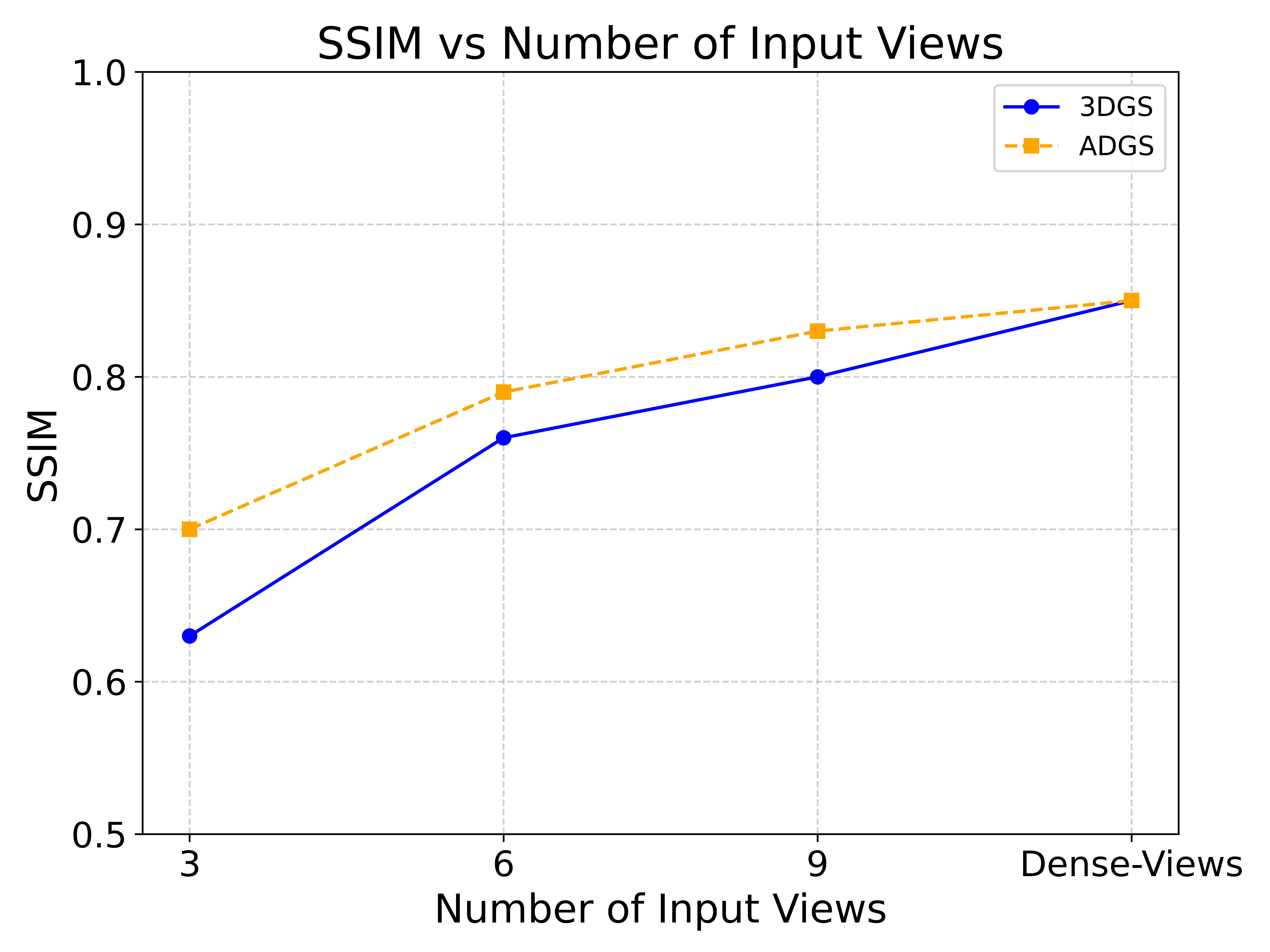}
    \caption{\textbf{SSIM vs number of input views for AD-GS and 3DGS on the entire LLFF dataset.} As the number of input views increases, our model performance converges to the dense-input 3DGS performance. Note that different scenes have different numbers of dense input views.}
    \label{fig:ssim-vs-views}
\end{figure}

 \begin{figure*}
    \centering
    \includegraphics[width=0.995\linewidth]{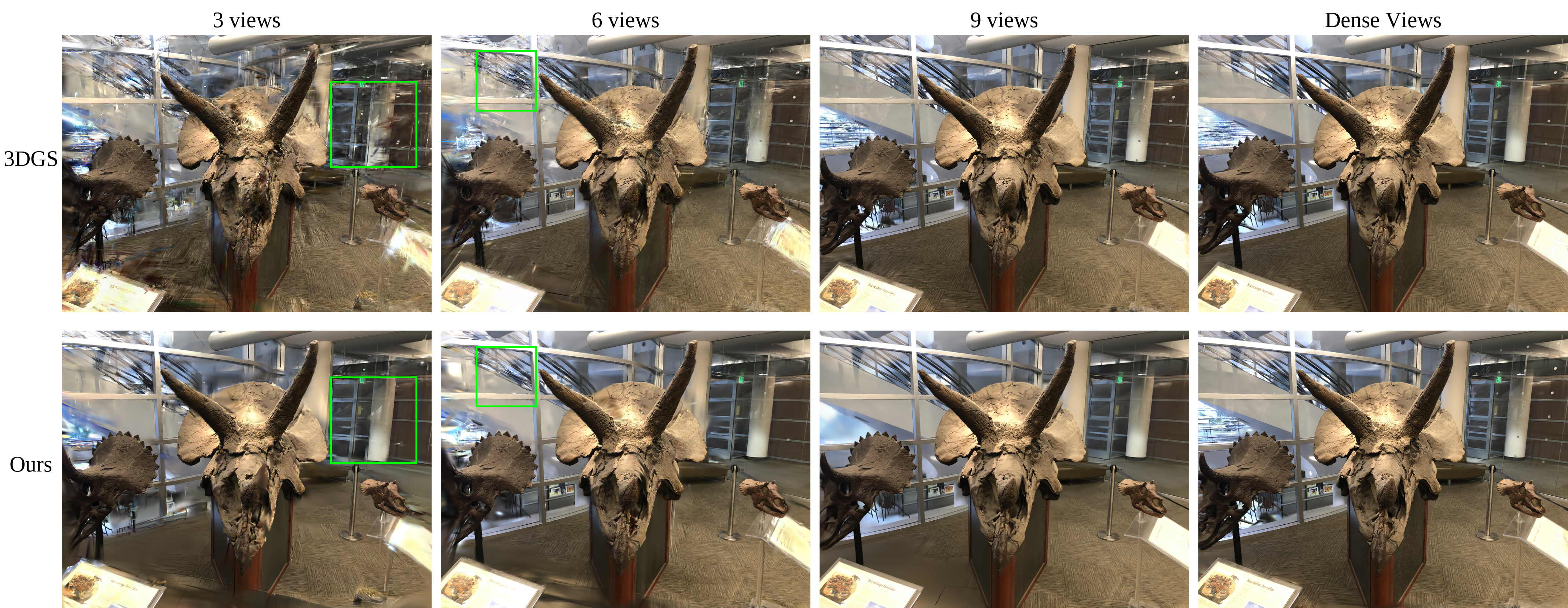}
    \caption{\textbf{Qualitative comparison on the LLFF dataset (Horns scene): AD-GS vs. 3DGS with increasing number of input views.}} AD-GS produces much cleaner and more stable reconstructions whereas 3DGS exhibits significant artifacts under sparse-view settings (e.g., 3 views) as seen in the green boxes. As the number of input views increases, the quality of AD-GS continues to improve and eventually converges to the performance of the dense-view 3DGS model.
    \label{fig:views-comparison}
\end{figure*}

\subsection{Results}

Tables \ref{tab:tanks_metrics}, \ref{tab:llff_metrics} and \ref{tab:m360-metrics} present the quantitative comparisons on all three datasets. We see that AD-GS consistently achieves superior performance on all metrics, for different number of input views and across all datasets. We see that AD-GS offers larger improvements when the number of input views is smaller. For instance, on the LLFF dataset with 3 input views, AD-GS improves SSIM from 0.674 (CoR-GS) to 0.699. Similarly, for LPIPS, our method achieves a notable gain over DropGaussian, reducing the score from 0.266 to 0.237, indicating better perceptual quality.

\cref{fig:results-m360-24,fig:results-m360-12,fig:results-llff-3,fig:llff-3-horns,fig:m360-24-room,fig:tanks-3-family,fig:llff-6-horns,fig:tanks-3-barn-churc} show the qualitative results across various scenes from all datasets. AD-GS is able to recover fine-scale textures and detailed structures that other methods tend to oversmooth or miss entirely. It maintains sharper object boundaries and better geometric fidelity, particularly noticeable in scenes with complex edges and discontinuities. Moreover, AD-GS suppresses floaters and noisy artifacts more effectively, resulting in better reconstructions. Its surface reconstructions also exhibit higher smoothness and realism, especially in areas with gradual depth variations. Thus our alternating training strategy offers benefits across a diverse range of sparse-view scenarios.

We also experiment with how our model performs as the number of input views reaches the dense setup.~\cref{fig:ssim-vs-views} illustrates that as the number of input views increases, AD-GS consistently outperforms 3DGS and converges more quickly towards the performance level of 3DGS in dense-view settings. We also see that there is no degradation in the performance of AD-GS with respect to 3DGS in the dense setting. This suggests a graceful performance convergence of AD-GS from the sparse to the dense view scenarios. 
 ~\cref{fig:views-comparison} shows superior reconstruction of AD-GS across sparse and dense view scenarios.

\subsection{Ablation Study}

We perform a detailed ablation study to evaluate the impact of each component in our AD-GS framework in  Table~\ref{tab:ablation_study_ssim_compact}. Our AD-GS pipeline uses both alternating densification (Alt-Dfn) and alternating losses (Alt-Loss). By alternating losses, we refer to our strategy of having different losses in different phases. To assess the individual contributions of these components, we compare against ablated variants that disable either component. In addition, we run another model to show the need for our densification approach. In particular, we employ only the combined loss $\mathcal{L_{\text{ds}}}$ throughout training without alternate densification. 

\begin{figure*}
    \centering
    \begin{subfigure}[b]{0.51\linewidth}
        \centering
        \includegraphics[width=\linewidth]{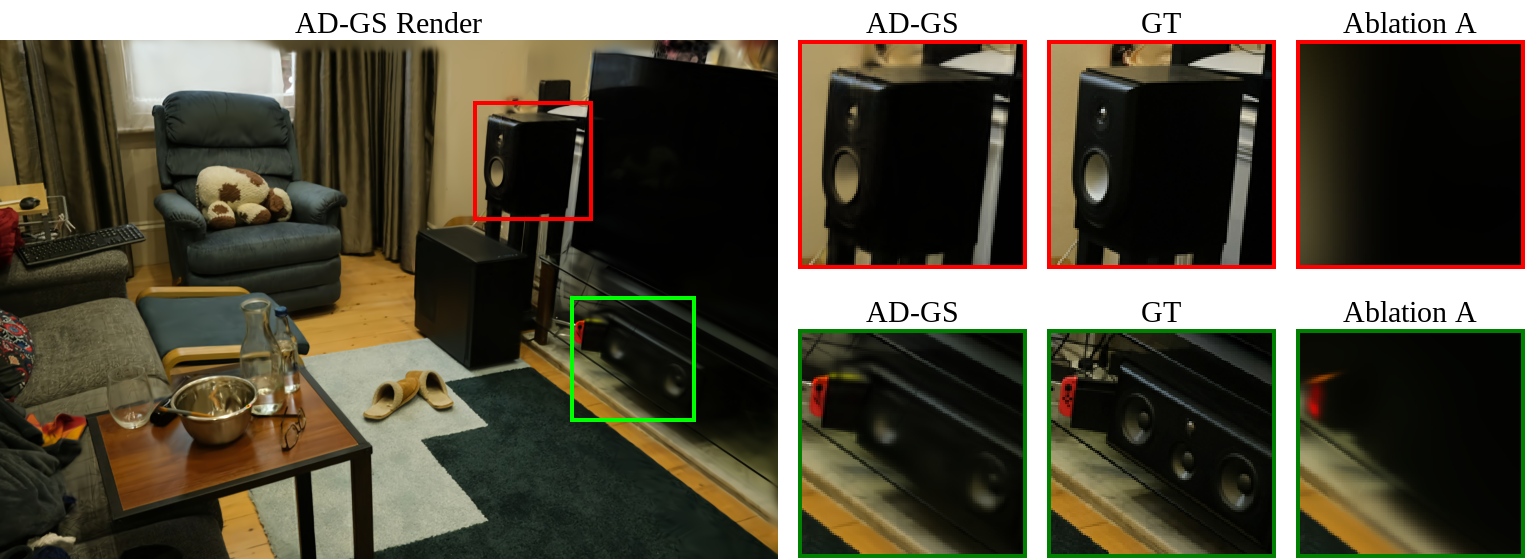}
        \caption{\textbf{Ablation A: AD-GS without alternating densification} struggles to learn finer details and instead predicts a smoother representation. }
        \label{fig:ab1}
    \end{subfigure}
    \hfill
    \begin{subfigure}[b]{0.47\linewidth}
        \centering
        \includegraphics[width=\linewidth]{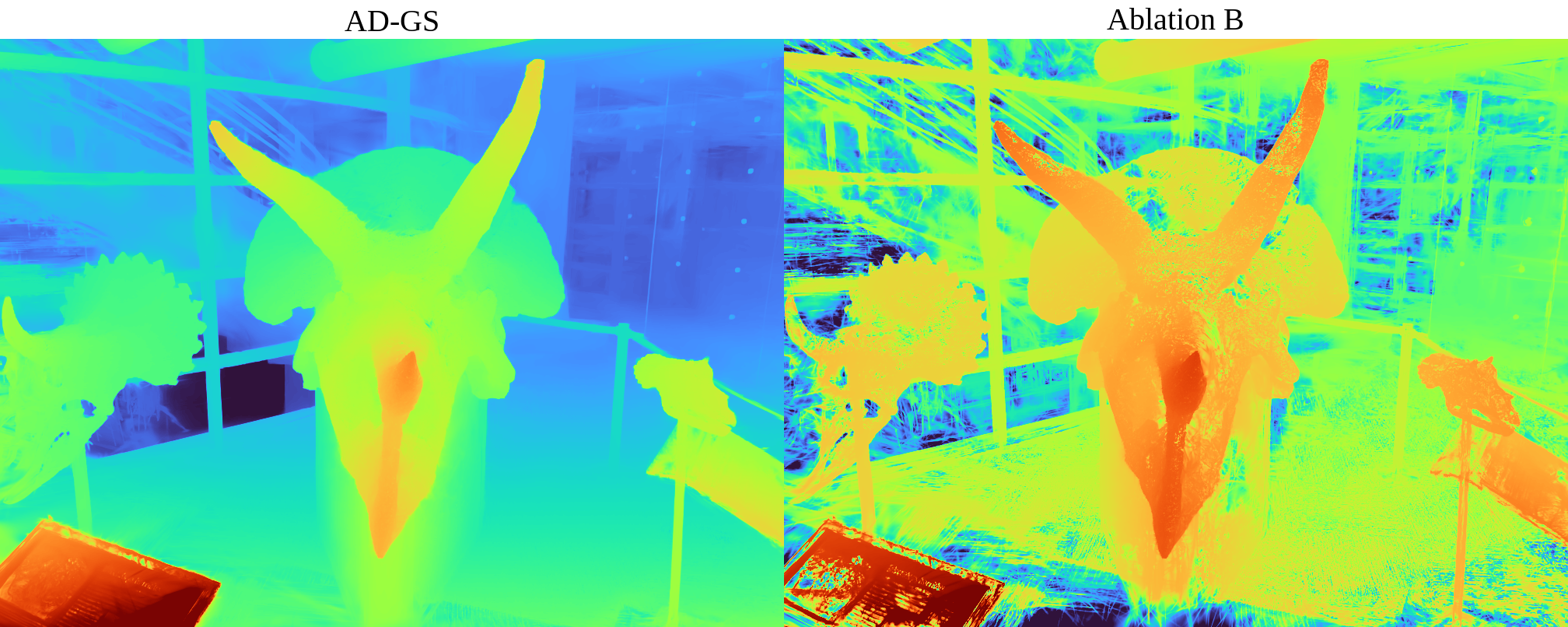}
        \caption{\textbf{Ablation B: AD-GS without alternating loss} leads to noisy depth maps with poor object structure.}
        \label{fig:ab2}
    \end{subfigure}
    \caption{Qualitative comparisons under different ablation settings}
    \label{fig:ablation_combined_1}
\end{figure*}

\begin{figure*}
    \centering
    \includegraphics[width=0.90\linewidth]{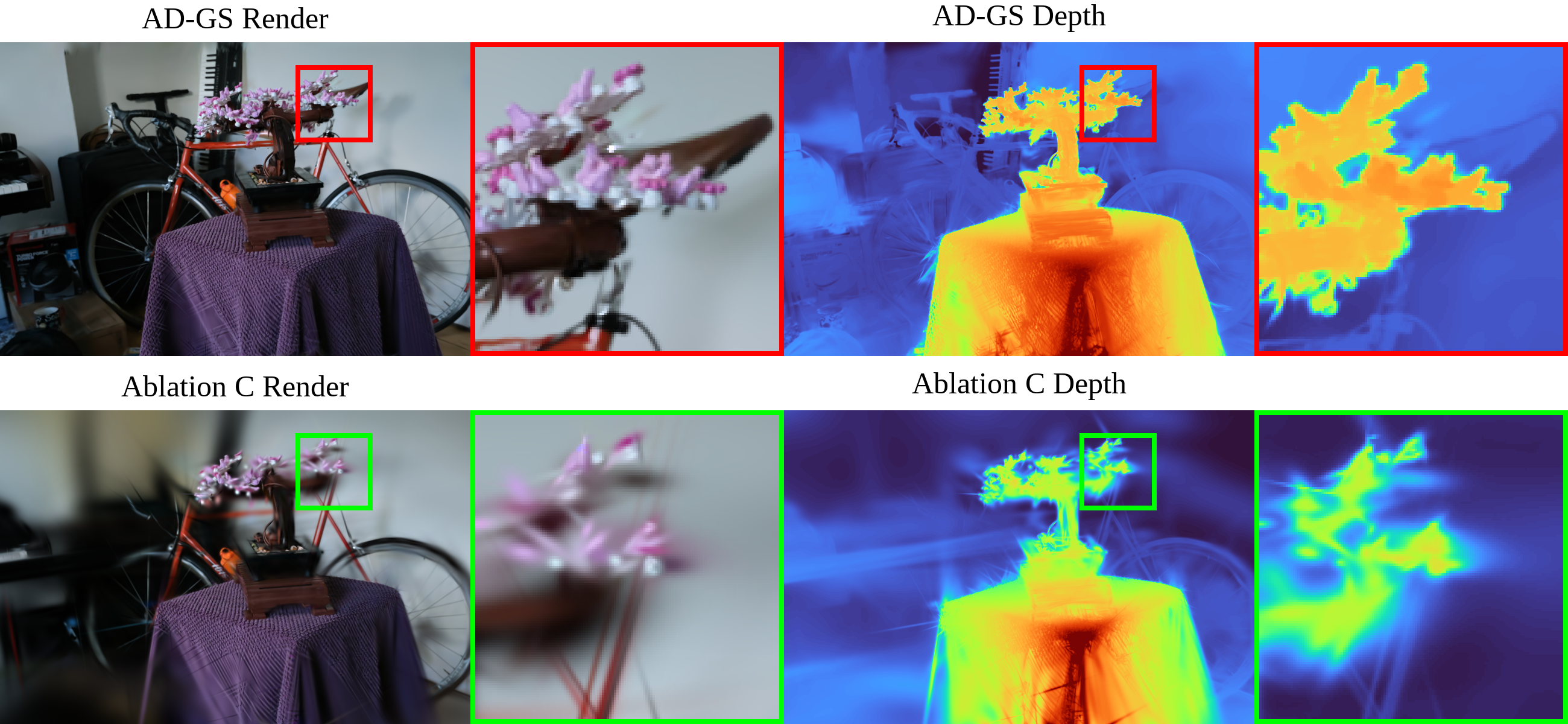}
    \caption{\textbf{Ablation C: AD-GS with full combined loss but no alternating densification} is unable to learn fine details and predicts a smooth output.}
    \label{fig:ablation_combined_2}
\end{figure*}

Our AD-GS model achieves the best performance across both LLFF (3/9-view) and Mip-NeRF360 (12-view) datasets. Disabling alternating densification (Row A) causes a noticeable drop in SSIM, indicating the importance of our alternate densification strategy. Similarly, removing alternating losses while retaining alternating densification (Row B) results in weaker SSIM, confirming the role of regularization in stabilizing geometry during sparse-view learning. We also evaluate a variant (Row C) trained with the full combined loss across all iterations without alternating losses. This results in excessive regularization, which hampers the model's ability to recover details, highlighting the necessity of phase-specific loss scheduling.
\cref{fig:ablation_combined_1} and \cref{fig:ablation_combined_2} illustrate the impact of ablations. 

\begin{table}[ht] 
\small
\caption{Ablation study on LLFF and Mip-NeRF360 datasets using SSIM.}
\label{tab:ablation_study_ssim_compact}
\begin{tabular}{l | c c | c}
\toprule
\textbf{Model / SSIM $\uparrow$} & \multicolumn{2}{c|}{\textbf{LLFF}} & \textbf{Mip-NeRF360} \\
& \textbf{3 views} & \textbf{9 views} & \textbf{12 views} \\
\midrule
Ours (AD-GS) & \cellcolor{best}0.699 & \cellcolor{best}0.830 & \cellcolor{best}0.593 \\
No Alt Dfn. (A) & \cellcolor{second}0.688 & \cellcolor{second}0.803 & \cellcolor{second}0.568 \\
No Alt Loss (B) & 0.644 & 0.796 & 0.542 \\
No Alt Dfn. + Comb. Loss (C) & \cellcolor{third}0.671 & \cellcolor{third}0.799 & \cellcolor{third}0.566 \\
\bottomrule
\end{tabular}
\end{table}

\subsection{Limitations}

While AD-GS demonstrates strong performance in sparse-view synthesis, it has a few limitations. Our framework requires two concurrently trained 3DGS models for pseudo-view consistency, which increases memory and computational overhead during training. Secondly, the pseudo-view consistency loss may be ineffective in situations where both models suffer from similar artifacts. Lastly, our approach tends to correct for the errors after they have been introduced by incorrect densification. It may be interesting to study geometry consistent densification and/or explore priors or learning based models to understand adaptive densification.

\section{Conclusion}

We present \textbf{AD-GS}, a novel framework designed for high-quality novel view synthesis under sparse input settings. By alternating between high and low densification phases, our approach establishes a self-correcting training mechanism that balances geometric regularization with detail recovery. Through this controlled training strategy, AD-GS mitigates overfitting and suppresses artifacts such as floaters, leading to improved scene reconstruction.
AD-GS achieves state-of-the-art performance on challenging benchmarks, producing significantly better reconstructions compared to existing baselines. In future, one could investigate whether adaptive durations of the two densification phases could further enhance performance.

\begin{acks}
    This work was supported in part by the Kotak IISc AI–ML Centre (KIAC).
\end{acks}

\clearpage
\newpage
\twocolumn
\appendix

{\centering\Huge \textbf{Supplementary Material}\par} 

\vspace{5mm}

\noindent This supplement provides additional details, experiments, and visualizations in support of the main paper. It includes:

\begin{enumerate}[label=\Alph*., leftmargin=2em]
    \item Video examples on all datasets.
    \item Details on databases, evaluation, and implementations.
    \item Additional ablation studies.
    \item Additional results.
    \item Extensive quantitative evaluation reports.
\end{enumerate}

\section{Video Examples on all Datasets}

We compare various models by rendering videos along a continuous trajectory. 
For the LLFF dataset, we render the videos along the spiral trajectory that is commonly used in the literature. 
For Mip-NeRF360, we render the videos along an elliptical trajectory, as used in the literature. 

The video comparisons are provided as part of the supplementary materials. 
Please refer to the included AD-GS videos folder to view qualitative comparisons across methods. 
It also contains the interactive HTML page that allows easy navigation and side-by-side visualization of the rendered videos.
The source code and video examples for our model can be found on our project page:
 \url{https://gurutvapatle.github.io/publications/2025/ADGS.html}.

\section{Database, Evaluation, and Implementation Details}

\subsection{Database Details}

We compare the performance of sparse input 3D Gaussian splatting models on the LLFF~\cite{mildenhall2019llff}, Mip-NeRF360~\cite{barron2022mipnerf360}, and Tanks \& Temples~\cite{knapitsch2017tanks} datasets with different sets of input views. 
For LLFF and Tanks \& Temples, we evaluate the models on 3, 6 and 9 input views. 
For Mip-NeRF360, we evaluate the models on 12 and 24 input views.

The LLFF dataset contains 8 forward facing unbounded scenes, each containing a variable number of frames at a spatial resolution of 4032$\times$3024. We downsample these images by a factor of 4 to obtain a resolution of 1008$\times$756. Following prior work \cite{zhang2024cor}, we use every 8\textsuperscript{th} frame for testing in each scene. We then uniformly sample \textit{n} training views from the remaining frames.

The Mip-NeRF360 dataset \cite{barron2022mipnerf360} contains seven publicly available 360 degree coverage unbounded scenes with high-resolution images. We downsample these images by a factor of 4, resulting in a range of final resolutions depending on the scene. Following prior work \cite{zhang2024cor}, we use every 8\textsuperscript{th} frame for testing and uniformly sample 12 or 24 training views from the remaining frames.

The Tanks and Temples dataset \cite{knapitsch2017tanks} contains 8 complex large-scale scenes with resolution of 960$\times$540. Following prior work \cite{zhang2024cor}, we use every 8\textsuperscript{th} frame for testing and uniformly sample 3, 6 or 9 training views from the rest.

\subsection{Evaluation Details}

In addition to using SSIM, PSNR, and LPIPS to compare model performances, to evaluate depth prediction quality, we use the depth 
Spearman rank order correlation coefficient (Depth SROCC) between the rendered depth and the pseudo ground truth depth, which computes the Pearson Linear Correlation coefficient (PLCC) between the ranks of the two depths.
Thus, the Depth SROCC evaluates the relative ordering of the objects in the scene.
Since ground-truth depth is unavailable, we train a vanilla 3DGS model with dense input views and use the rendered depth on the test views as the pseudo ground truth. 

We also compute video metrics between the rendered videos for spiral or elliptical trajectories and the pseudo ground truth video, where the pseudo ground truth video is also obtained using the 3DGS model trained with dense input views. In particular,  we render spiral camera trajectories for each scene of LLFF dataset and ellipse camera trajectories for each scene of Mip-NeRF360 dataset, starting from the given COLMAP poses. 
We compute frame-wise SSIM, PSNR and LPIPS of the rendered videos with respect to the pseudo-ground truth video, to capture overall visual consistency and fidelity.

\subsection{Implementation Details}

Our training pipeline consists of a warm-up phase and an alternating densification phase. During the warm-up phase, we follow a similar training strategy as employed in 3DGS, FSGS and CoR-GS, using photometric supervision to establish an initial scene geometry. 
In the alternating stage, we switch between two optimization phases every $N_L$ and $N_H$  iterations. 

The loss weights are chosen to ensure that all terms contribute with comparable magnitude. All hyperparameters are kept fixed across all datasets and scenes and no tuning is performed per scene. The complete set of parameters is summarized in Table \ref{tab:training_config}.





\begin{table}[H]
\small
\caption{Training configuration and hyperparameters used in our method (AD-GS). All values are fixed across datasets and scenes.}
\label{tab:training_config}
\begin{tabular}{l | l}
\toprule
\textbf{Component} & \textbf{Value / Description} \\
\midrule
Warm-up phase & 1500 iterations \\
Total training iterations & 10,000 \\
Alternation phase lengths & $N_L = 100$, $N_H = 100$ \\
\midrule
Low densification thresholds & $\epsilon_L = 0.1$, $\tau_L = 0.0005$ \\
High densification thresholds & $\epsilon_\alpha = 0.005$, $\tau_{\text{pos}} = 0.0002$ \\
\midrule
Combined loss weights & $\lambda_1 = 1$, $\lambda_2 = 1$, $\lambda_3 = 1$ \\
Edge-aware smoothness loss &  \( \lambda_{\text{r}} = 0.001 \), $\omega_1 = 0.01$, $\omega_2 = 0.05 $ \\
\midrule
GPU used & NVIDIA RTX 3090 (24GB) \\
Baselines & Official codebases and configs \\
\bottomrule
\end{tabular}
\end{table}

\subsection{Memory Usage and Training Time}

Table~\ref{tab:memory_time} summarizes peak GPU memory consumption and per-scene training time for LLFF (3-view) on an NVIDIA RTX 3090 GPU.
The training time overhead is moderate, justified by the improved quality. AD-GS takes about 15 minutes per scene, similar to CoR-GS, and slightly longer than 3DGS and FSGS. 
During inference of AD-GS, only a single model is used, so runtime latency is similar to 3DGS. 

\begin{table}[H]
\centering
\caption{Peak GPU memory usage (\%) and training time (minutes) for LLFF (3-view) on NVIDIA RTX 3090.}
\label{tab:memory_time}
\begin{tabular}{l | c | c }
\toprule
{Method } & {Peak GPU Memory} & {Training Time } \\
(on LLFF 3-view) & Usage (\%) & (minutes)\\
\midrule
AD-GS (Ours) & 89\% & $\sim 15$ \\
3DGS        & 72\% & $\sim 10$ \\
FSGS        & 78\% & $\sim 12$ \\
CoR-GS      & 88\% & $\sim 15$ \\
DropGaussian      & 75\% & $\sim 10$ \\
\bottomrule
\end{tabular}
\end{table} 

\section{Additional Ablation Studies}
\label{sec:additional_ablation}

\subsection{Extended Ablation Study}

We complement the main ablation study with additional experiments to analyze the impact of removing individual loss components.
Removing the pseudo-view consistency loss (Row D) from Equation (9) leads to a noticeable drop in SSIM. 
Similarly, a degradation in performance occurs when the edge-aware depth smoothness loss (Row E) is removed from Equation (7), even if the range loss is retained. Furthermore, eliminating both the range loss and the edge-aware depth smoothness term (Row F) from Equation (7) leads to poorer performance due to depth instability. The results of these ablations are summarized in Table~\ref{tab:additional_ablation_study_ssim}.
These results highlight the importance of each loss component in ensuring stable and accurate reconstructions in sparse-view settings.

\begin{table}[H] 
\small
\caption{Additional ablation study using SSIM.}
\label{tab:additional_ablation_study_ssim}
\begin{tabular}{l | c c | c}
\toprule
\textbf{Model / SSIM $\uparrow$} & \multicolumn{2}{c|}{\textbf{LLFF}} & \textbf{Mip-NeRF360} \\
& \textbf{3 views} & \textbf{9 views} & \textbf{12 views} \\
\midrule
\midrule
Ours (AD-GS) & \cellcolor{best}0.699 & \cellcolor{best}0.830 & \cellcolor{best}0.593 \\
\midrule
No Alt Dfn. (A) & \cellcolor{second}0.688 & \cellcolor{second}0.803 & \cellcolor{second}0.568 \\
\midrule
No Alt Loss (B) & 0.644 & 0.796 & 0.542 \\
\midrule
No Alt Dfn. + Comb. Loss (C) & 0.671 & 0.799 & 0.566 \\
\midrule
w/o Pseudo-View& 0.658 & 0.799 & 0.549 \\
 Consistency Loss (D)  & & & \\
\midrule
w/o Edge-Aware Depth & \cellcolor{third}0.678 & \cellcolor{third}0.803 & \cellcolor{third}0.567 \\
Smoothness Loss (E) & \cellcolor{third} & \cellcolor{third} & \cellcolor{third} \\
\midrule
w/o Edge-Aware Depth & 0.672 & 0.800 & 0.560 \\
Smoothness Loss & & & \\
+  w/o Depth Range Loss (F) & & & \\
\bottomrule
\end{tabular}
\end{table}

\subsection{Phase Length Sensitivity}

We evaluate how changing the length of each alternation phase affects performance. Results show that longer phases can slightly degrade quality due to uncontrolled Gaussian growth as shown in Table \ref{tab:iteration_vs_ssim}.

\begin{table}[H]
\centering
\caption{Effect of training phase length on SSIM for LLFF (3-view). Longer phases slightly reduce quality.}
\label{tab:iteration_vs_ssim}
\begin{tabular}{c | c}
\toprule
{Phase Length (Iterations)} & {SSIM $\uparrow$ (LLFF 3-view)} \\
\midrule
100 & 0.699 \\
300 & 0.691 \\
500 & 0.689 \\
1000 & 0.688 \\
\bottomrule
\end{tabular}
\end{table}

\section{Additional Results}

We present comparisons with respect to Depth SROCC, video metrics and model size in ~\cref{tab:metrics_step_3,tab:metrics_step_6,tab:metrics_step_9} on the LLFF (for 3, 6 and 9 views) and ~\cref{tab:metrics_step_12,tab:metrics_step_24} on Mip-NeRF360 (for 12 and 24 views) datasets. We provide inferences on each of these measures as follows:

Methods like CoR-GS tend to employ strong smoothing while DropGaussian reduces model complexity by randomly removing Gaussians during training. These strategies result in smaller model sizes, particularly when the number of input views is limited (e.g., 3 or 6 views on the LLFF dataset). This leads to a limitation in their ability to capture finer details which is evident in Video SSIM comparisons. However, as the number of views increases (e.g., 9 on LLFF or 24 on Mip-NeRF360), our method, AD-GS, achieves the smallest model size among all approaches despite offering good reconstruction quality as seen in the video metrics. This indicates that AD-GS scales more effectively with increased supervision. On the other hand, 3DGS tends to over-densify, resulting in large model sizes regardless of the dataset or input configuration.

Our model consistently outperforms all other baselines in terms of Depth SROCC, demonstrating superior correlation with pseudo-ground truth depth rankings across diverse scenes.
Notably, our model maintains its high SROCC scores across both the LLFF and Mip-NeRF360 datasets, which may be attributed to our alternative densification strategy. We also note that we see improvements in the video metrics that are consistent with the improvements seen for image metrics presented in the main paper. 


\section{Extensive quantitative evaluation reports}

We provide comprehensive scene-wise quantitative results for all scenes in both the LLFF and Mip-NeRF360 datasets. These results cover all evaluation metrics reported in the paper, enabling a detailed comparison across different methods and input settings. Please refer to the tables below for a complete breakdown of the per-scene performance.

\begin{table*}
\centering
\caption{Quantitative Results on LLFF dataset with 3 input views}
\begin{tabular}{lccccc}
\hline
\textbf{Method} & \textbf{Video SSIM} & \textbf{Video PSNR} & \textbf{Video LPIPS} & \textbf{SROCC} & \textbf{Model Size (MB)} \\
\hline
3DGS     & 0.650 & 18.82 & 0.260 & 0.745 & 62.75 \\
FSGS     & 0.675 & 19.50 & 0.252 & 0.796 & 49.32 \\
CoR-GS    & 0.701 & 19.74 & 0.245 & 0.807 & 22.18 \\
DropGaussian   & 0.699 & 19.78 & 0.242 & 0.780 & 22.89 \\
ADGS     & 0.728 & 20.20 & 0.212 & 0.812 & 42.01 \\
\hline
\end{tabular}
\label{tab:metrics_step_3}
\end{table*}

\begin{table*}
\centering
\caption{Quantitative Results on LLFF dataset with 6 input views}
\begin{tabular}{lccccc}
\hline
\textbf{Method} & \textbf{Video SSIM} & \textbf{Video PSNR} & \textbf{Video LPIPS} & \textbf{SROCC} & \textbf{Model Size (MB)} \\
\hline
3DGS     & 0.810 & 23.15 & 0.149 & 0.880 & 114.44 \\
FSGS     & 0.830 & 24.21 & 0.140 & 0.837 & 62.71 \\
CoR-GS    & 0.838 & 24.43 & 0.142 & 0.888 & 50.54 \\
DropGaussian   & 0.843 & 24.54 & 0.134 & 0.878 & 51.07 \\
ADGS     & 0.852 & 24.71 & 0.123 & 0.889 & 59.16 \\
\hline
\end{tabular}
\label{tab:metrics_step_6}
\end{table*}

\begin{table*}
\centering
\caption{Quantitative Results on LLFF dataset with 9 input views}
\begin{tabular}{lccccc}
\hline
\textbf{Method} & \textbf{Video SSIM} & \textbf{Video PSNR} & \textbf{Video LPIPS} & \textbf{SROCC} & \textbf{Model Size (MB)} \\
\hline
3DGS     & 0.858 & 25.34 & 0.118 & 0.892 & 185.58 \\
FSGS     & 0.875 & 26.07 & 0.109 & 0.847 & 109.02 \\
CoR-GS    & 0.878 & 26.38 & 0.111 & 0.906 & 108.49 \\
DropGaussian   & 0.884 & 26.46 & 0.103 & 0.905 & 107.35 \\
ADGS     & 0.891 & 26.76 & 0.100 & 0.916 & 89.98 \\
\hline
\end{tabular}
\label{tab:metrics_step_9}
\end{table*}

\begin{table*}
\centering
\caption{Quantitative Results on Mip-NeRF360 dataset with 12 input views.}
\begin{tabular}{lccccc}
\hline
\textbf{Method} & \textbf{Video SSIM} & \textbf{Video PSNR} & \textbf{Video LPIPS} & \textbf{SROCC} & \textbf{Model Size (MB)} \\
\hline
3DGS     & 0.528 & 18.39 & 0.379 & 0.537 & 326.12 \\
FSGS     & 0.585 & 19.31 & 0.378 & 0.788 & 101.49 \\
CoR-GS    & 0.631 & 20.48 & 0.369 & 0.692 & 101.84 \\
DropGaussian   & 0.625 & 20.19 & 0.367 & 0.641 & 101.13 \\
ADGS (30k) & 0.632 & 20.38 & 0.344 & 0.806 & 111.75 \\
ADGS (10k) & 0.646 & 20.75 & 0.343 & 0.794 & 109.62 \\
\hline
\end{tabular}
\label{tab:metrics_step_12}
\end{table*}

\begin{table*}
\centering
\caption{Quantitative Results on Mip-NeRF360 dataset with 24 input views.}
\begin{tabular}{lccccc}
\hline
\textbf{Method} & \textbf{Video SSIM} & \textbf{Video PSNR} & \textbf{Video LPIPS} & \textbf{SROCC} & \textbf{Model Size (MB)} \\
\hline
3DGS       & 0.741 & 23.85 & 0.232 & 0.800 & 595.29 \\
FSGS       & 0.769 & 24.95 & 0.238 & 0.893 & 316.54 \\
CoR-GS      & 0.788 & 25.37 & 0.227 & 0.814 & 332.38 \\
DropGaussian     & 0.794 & 25.55 & 0.225 & 0.825 & 325.96 \\
ADGS (30k)   & 0.807 & 25.69 & 0.196 & 0.907 & 206.05 \\
ADGS (10k)   & 0.814 & 25.93 & 0.198 & 0.906 & 231.44 \\
\hline
\end{tabular}
\label{tab:metrics_step_24}
\end{table*}

\begin{table*}[ht]
\centering
\caption{Scene-wise Quantitative Results on LLFF dataset with 3 input views.}
\vspace{0.5em}
\begin{tabular}{llccccccccc}
\toprule
\textbf{Method} & \textbf{Metric} & \textbf{fern} & \textbf{flower} & \textbf{fortress} & \textbf{horns} & \textbf{leaves} & \textbf{orchids} & \textbf{room} & \textbf{trex} & \textbf{Average} \\
\midrule

{3DGS} 
& SSIM & 0.652 & 0.589 & 0.647 & 0.599 & 0.514 & 0.421 & 0.847 & 0.791 & 0.632 \\
& PSNR & 19.84 & 19.20 & 21.25 & 17.59 & 15.74 & 14.85 & 21.61 & 21.08 & 18.90 \\
& LPIPS & 0.262 & 0.316 & 0.263 & 0.323 & 0.273 & 0.335 & 0.184 & 0.197 & 0.269 \\
& Video SSIM & 0.663 & 0.577 & 0.683 & 0.614 & 0.551 & 0.440 & 0.841 & 0.831 & 0.650 \\
& Video PSNR & 19.90 & 17.14 & 21.63 & 17.76 & 16.12 & 14.84 & 21.14 & 22.03 & 18.82 \\
& Video LPIPS & 0.256 & 0.326 & 0.236 & 0.318 & 0.274 & 0.337 & 0.187 & 0.147 & 0.260 \\
& Depth SROCC & 0.800 & 0.658 & 0.695 & 0.777 & 0.726 & 0.652 & 0.816 & 0.836 & 0.745 \\
& Model Size &  73.83 & 52.92 & 35.59 & 60.30 & 154.74 & 58.04 & 25.83 & 40.76 & 62.75 \\

\midrule

{FSGS} 
& SSIM & 0.665 & 0.590 & 0.687 & 0.663 & 0.534 & 0.445 & 0.860 & 0.796 & 0.655 \\
& PSNR & 19.99 & 19.17 & 22.50 & 19.36 & 16.53 & 15.61 & 21.58 & 21.26 & 19.50 \\
& LPIPS & 0.270 & 0.332 & 0.249 & 0.305 & 0.276 & 0.329 & 0.182 & 0.211 & 0.269 \\
& Video SSIM & 0.676 & 0.582 & 0.727 & 0.685 & 0.567 & 0.470 & 0.858 & 0.836 & 0.675 \\
& Video PSNR & 19.97 & 17.74 & 23.38 & 19.21 & 16.82 & 15.66 & 20.99 & 22.25 & 19.50 \\
& Video LPIPS & 0.259 & 0.336 & 0.211 & 0.280 & 0.278 & 0.329 & 0.166 & 0.154 & 0.252 \\
& Depth SROCC & 0.722 & 0.800 & 0.897 & 0.910 & 0.703 & 0.781 & 0.764 & 0.793 & 0.796 \\
& Model Size &  31.08 & 56.60 & 20.22 & 32.22 & 163.34 & 38.59 & 15.57 & 36.98 & 49.33 \\

\midrule

{CoR-GS} 
& SSIM & 0.693 & 0.642 & 0.729 & 0.671 & 0.518 & 0.456 & 0.867 & 0.816 & 0.674 \\
& PSNR & 20.74 & 19.76 & 23.03 & 18.98 & 15.81 & 15.24 & 21.57 & 21.62 & 19.59 \\
& LPIPS & 0.267 & 0.319 & 0.241 & 0.328 & 0.314 & 0.331 & 0.172 & 0.193 & 0.271 \\
& Video SSIM & 0.706 & 0.650 & 0.774 & 0.700 & 0.567 & 0.488 & 0.869 & 0.857 & 0.701 \\
& Video PSNR & 20.82 & 18.45 & 24.01 & 18.78 & 16.52 & 15.50 & 21.10 & 22.74 & 19.74 \\
& Video LPIPS & 0.249 & 0.307 & 0.202 & 0.289 & 0.301 & 0.334 & 0.150 & 0.129 & 0.245 \\
& Depth SROCC & 0.830 & 0.821 & 0.842 & 0.855 & 0.712 & 0.733 & 0.779 & 0.883 & 0.807 \\
& Model Size &  24.49 & 20.14 & 14.23 & 19.61 & 50.94 & 21.94 & 9.68 & 16.42 & 22.18 \\

\midrule

{DropGaussian} 
& SSIM & 0.689 & 0.646 & 0.709 & 0.657 & 0.547 & 0.462 & 0.863 & 0.819 & 0.674 \\
& PSNR & 20.68 & 20.28 & 22.57 & 18.39 & 16.45 & 15.63 & 21.76 & 21.95 & 19.72 \\
& LPIPS & 0.265 & 0.315 & 0.252 & 0.330 & 0.276 & 0.322 & 0.175 & 0.192 & 0.266 \\
& Video SSIM & 0.707 & 0.653 & 0.751 & 0.681 & 0.596 & 0.486 & 0.861 & 0.861 & 0.699 \\
& Video PSNR & 21.08 & 18.69 & 23.16 & 18.22 & 17.00 & 15.73 & 21.23 & 23.15 & 19.78 \\
& Video LPIPS & 0.249 & 0.301 & 0.212 & 0.300 & 0.263 & 0.322 & 0.159 & 0.128 & 0.242 \\
& Depth SROCC & 0.808 & 0.784 & 0.716 & 0.834 & 0.755 & 0.688 & 0.763 & 0.892 & 0.780 \\
& Model Size & 24.64 & 19.09 & 15.93 & 20.20 & 50.32 & 23.43 & 11.62 & 17.90 & 22.89 \\

\midrule

{AD-GS} 
& SSIM & 0.705 & 0.648 & 0.772 & 0.714 & 0.578 & 0.474 & 0.879 & 0.823 & 0.699 \\
& PSNR & 20.82 & 19.86 & 23.87 & 19.79 & 16.64 & 15.45 & 22.39 & 21.68 & 20.06 \\
& LPIPS & 0.243 & 0.288 & 0.207 & 0.274 & 0.251 & 0.304 & 0.155 & 0.178 & 0.237 \\
& Video SSIM & 0.716 & 0.657 & 0.813 & 0.745 & 0.640 & 0.507 & 0.880 & 0.868 & 0.728 \\
& Video PSNR & 20.79 & 18.49 & 25.30 & 19.34 & 17.35 & 15.58 & 21.83 & 22.94 & 20.20 \\
& Video LPIPS & 0.225 & 0.274 & 0.164 & 0.236 & 0.239 & 0.315 & 0.133 & 0.112 & 0.212 \\
& Depth SROCC & 0.834 & 0.853 & 0.879 & 0.864 & 0.670 & 0.699 & 0.809 & 0.890 & 0.812 \\
& Model Size & 37.96 & 30.28 & 20.53 & 30.52 & 89.98 & 90.38 & 11.86 & 24.60 & 42.01 \\

\bottomrule
\end{tabular}
\end{table*}

\begin{table*}[ht]
\centering
\caption{Scene-wise Quantitative Results on LLFF dataset with 6 input views.}
\vspace{0.5em}
\begin{tabular}{llcccccccccc}
\toprule
\textbf{Method} & \textbf{Metric} & \textbf{fern} & \textbf{flower} & \textbf{fortress} & \textbf{horns} & \textbf{leaves} & \textbf{orchids} & \textbf{room} & \textbf{trex} & \textbf{{Average}} \\
\midrule
{3DGS} & SSIM & 0.778 & 0.766 & 0.794 & 0.796 & 0.628 & 0.510 & 0.931 & 0.857 & 0.757 \\
& PSNR & 22.78 & 23.89 & 24.69 & 23.31 & 17.71 & 16.52 & 28.33 & 23.34 & 22.57 \\
& LPIPS & 0.173 & 0.181 & 0.171 & 0.184 & 0.207 & 0.287 & 0.100 & 0.154 & 0.182 \\
& Video SSIM & 0.820 & 0.837 & 0.849 & 0.818 & 0.725 & 0.613 & 0.922 & 0.897 & 0.810 \\
& Video PSNR & 24.11 & 25.73 & 24.99 & 23.07 & 19.21 & 17.94 & 25.46 & 24.66 & 23.15 \\
& Video LPIPS & 0.127 & 0.137 & 0.140 & 0.172 & 0.176 & 0.227 & 0.114 & 0.102 & 0.149 \\
& Depth SROCC & 0.910 & 0.902 & 0.878 & 0.933 & 0.821 & 0.805 & 0.862 & 0.931 & 0.880 \\
& Model Size &  108.77 & 72.89 & 76.97 & 124.25 & 254.91 & 120.21 & 56.58 & 100.96 & 114.44 \\

\midrule
{FSGS} & SSIM & 0.792 & 0.776 & 0.840 & 0.819 & 0.625 & 0.548 & 0.935 & 0.874 & 0.776 \\
& PSNR & 23.20 & 24.12 & 27.76 & 23.61 & 18.10 & 17.21 & 28.15 & 23.85 & 23.25 \\
& LPIPS & 0.186 & 0.199 & 0.149 & 0.193 & 0.224 & 0.270 & 0.109 & 0.153 & 0.185 \\
& Video SSIM & 0.833 & 0.848 & 0.895 & 0.855 & 0.714 & 0.647 & 0.935 & 0.918 & 0.831 \\
& Video PSNR & 24.69 & 25.98 & 29.62 & 23.31 & 19.36 & 18.68 & 26.32 & 25.74 & 24.21 \\
& Video LPIPS & 0.137 & 0.136 & 0.108 & 0.148 & 0.192 & 0.220 & 0.097 & 0.085 & 0.140 \\
& Depth SROCC & 0.777 & 0.869 & 0.901 & 0.947 & 0.750 & 0.850 & 0.789 & 0.815 & 0.837 \\
& Model Size & 47.31 & 33.75 & 75.19 & 70.02 & 124.27 & 56.72 & 36.51 & 57.95 & 62.72 \\

\midrule
{CoR-GS} & SSIM & 0.797 & 0.783 & 0.835 & 0.825 & 0.623 & 0.554 & 0.940 & 0.883 & 0.780 \\
& PSNR & 23.44 & 24.30 & 26.94 & 24.05 & 17.71 & 17.24 & 28.68 & 24.30 & 23.33 \\
& LPIPS & 0.182 & 0.205 & 0.153 & 0.196 & 0.250 & 0.275 & 0.096 & 0.139 & 0.187 \\
& Video SSIM & 0.836 & 0.863 & 0.892 & 0.862 & 0.730 & 0.657 & 0.939 & 0.926 & 0.838 \\
& Video PSNR & 24.91 & 26.44 & 29.04 & 23.53 & 19.45 & 18.81 & 26.84 & 26.39 & 24.43 \\
& Video LPIPS & 0.137 & 0.137 & 0.110 & 0.145 & 0.213 & 0.233 & 0.088 & 0.073 & 0.142 \\
& Depth SROCC & 0.905 & 0.917 & 0.889 & 0.951 & 0.797 & 0.863 & 0.869 & 0.918 & 0.889 \\
& Model Size & 43.02 & 29.24 & 64.00 & 67.03 & 74.08 & 45.67 & 37.19 & 44.10 & 50.54 \\

\midrule
{DropGaussian} & SSIM & 0.805 & 0.792 & 0.846 & 0.824 & 0.640 & 0.562 & 0.939 & 0.881 & 0.786 \\
& PSNR & 23.74 & 24.73 & 27.77 & 23.74 & 18.11 & 17.31 & 28.76 & 24.04 & 23.52 \\
& LPIPS & 0.181 & 0.199 & 0.156 & 0.193 & 0.223 & 0.264 & 0.096 & 0.144 & 0.182 \\
& Video SSIM & 0.844 & 0.869 & 0.902 & 0.860 & 0.741 & 0.662 & 0.938 & 0.925 & 0.843 \\
& Video PSNR & 25.25 & 26.87 & 29.73 & 23.06 & 19.62 & 18.73 & 26.82 & 26.24 & 24.54 \\
& Video LPIPS & 0.135 & 0.129 & 0.106 & 0.142 & 0.185 & 0.214 & 0.087 & 0.074 & 0.134 \\
& Depth SROCC & 0.905 & 0.910 & 0.885 & 0.938 & 0.802 & 0.791 & 0.857 & 0.935 & 0.878 \\
& Model Size &  41.68 & 30.61 & 62.15 & 66.61 & 77.28 & 49.75 & 38.18 & 42.27 & 51.07 \\

\midrule
{AD-GS} & SSIM & 0.801 & 0.808 & 0.851 & 0.836 & 0.647 & 0.568 & 0.944 & 0.887 & 0.793 \\
& PSNR & 23.33 & 24.85 & 27.48 & 24.06 & 18.27 & 17.10 & 28.89 & 24.32 & 23.54 \\
& LPIPS & 0.178 & 0.171 & 0.139 & 0.178 & 0.222 & 0.249 & 0.090 & 0.134 & 0.170 \\
& Video SSIM & 0.844 & 0.886 & 0.905 & 0.875 & 0.755 & 0.676 & 0.946 & 0.930 & 0.852 \\
& Video PSNR & 25.00 & 27.05 & 29.66 & 23.70 & 19.98 & 18.82 & 27.08 & 26.41 & 24.71 \\
& Video LPIPS & 0.127 & 0.104 & 0.095 & 0.126 & 0.183 & 0.207 & 0.077 & 0.065 & 0.123 \\
& Depth SROCC & 0.905 & 0.933 & 0.908 & 0.958 & 0.745 & 0.875 & 0.862 & 0.929 & 0.889 \\
& Model Size &  51.89 & 45.46 & 51.89 & 62.12 & 120.55 & 70.03 & 27.10 & 44.24 & 59.16 \\

\bottomrule
\end{tabular}
\end{table*}

\begin{table*}[ht]
\centering
\caption{Scene-wise Quantitative Results on LLFF dataset with 9 input views.}
\vspace{0.5em}
\begin{tabular}{llcccccccccc}
\toprule
\textbf{Method} & \textbf{Metric} & \textbf{fern} & \textbf{flower} & \textbf{fortress} & \textbf{horns} & \textbf{leaves} & \textbf{orchids} & \textbf{room} & \textbf{trex} & \textbf{Average} \\
\midrule
{3DGS} & SSIM & 0.831 & 0.814 & 0.783 & 0.863 & 0.691 & 0.565 & 0.948 & 0.899 & 0.799 \\
& PSNR & 24.37 & 25.56 & 25.60 & 25.12 & 19.02 & 17.64 & 29.76 & 25.09 & 24.02 \\
& LPIPS & 0.138 & 0.144 & 0.167 & 0.125 & 0.177 & 0.254 & 0.084 & 0.119 & 0.151 \\
& Video SSIM & 0.885 & 0.888 & 0.857 & 0.880 & 0.796 & 0.673 & 0.947 & 0.935 & 0.858 \\
& Video PSNR & 26.41 & 27.88 & 27.24 & 24.96 & 20.93 & 19.19 & 28.71 & 27.39 & 25.34 \\
& Video LPIPS & 0.086 & 0.101 & 0.136 & 0.131 & 0.137 & 0.193 & 0.090 & 0.071 & 0.118 \\
& Depth SROCC & 0.921 & 0.906 & 0.887 & 0.955 & 0.793 & 0.833 & 0.909 & 0.934 & 0.892 \\
& Model Size &  198.70 & 121.12 & 160.41 & 218.54 & 359.51 & 197.47 & 96.27 & 132.63 & 185.58 \\

\midrule
{FSGS} & SSIM & 0.846 & 0.822 & 0.823 & 0.872 & 0.693 & 0.603 & 0.951 & 0.907 & 0.815 \\
& PSNR & 24.83 & 25.74 & 27.69 & 24.97 & 19.37 & 18.31 & 29.44 & 26.17 & 24.56 \\
& LPIPS & 0.137 & 0.159 & 0.145 & 0.137 & 0.186 & 0.235 & 0.089 & 0.122 & 0.151 \\
& Video SSIM & 0.891 & 0.902 & 0.896 & 0.901 & 0.797 & 0.710 & 0.955 & 0.944 & 0.875 \\
& Video PSNR & 26.71 & 28.19 & 30.39 & 24.95 & 21.25 & 20.09 & 28.98 & 27.98 & 26.07 \\
& Video LPIPS & 0.092 & 0.093 & 0.103 & 0.111 & 0.143 & 0.188 & 0.077 & 0.061 & 0.109 \\
& Depth SROCC & 0.790 & 0.893 & 0.903 & 0.954 & 0.757 & 0.867 & 0.791 & 0.820 & 0.847 \\
& Model Size   & 95.21 & 62.30 & 151.97 & 179.53 & 130.99 & 91.93 & 74.02 & 86.22 & 109.02 \\

\midrule
{CoR-GS} & SSIM & 0.847 & 0.825 & 0.813 & 0.872 & 0.699 & 0.610 & 0.954 & 0.914 & 0.817 \\
& PSNR & 24.88 & 25.84 & 27.65 & 25.08 & 19.29 & 18.57 & 30.02 & 26.63 & 24.75 \\
& LPIPS & 0.137 & 0.164 & 0.150 & 0.136 & 0.196 & 0.240 & 0.079 & 0.110 & 0.152 \\
& Video SSIM & 0.895 & 0.910 & 0.891 & 0.899 & 0.808 & 0.719 & 0.958 & 0.949 & 0.879 \\
& Video PSNR & 26.99 & 28.92 & 30.45 & 24.73 & 21.55 & 20.37 & 29.57 & 28.42 & 26.38 \\
& Video LPIPS & 0.093 & 0.093 & 0.106 & 0.114 & 0.155 & 0.202 & 0.071 & 0.054 & 0.111 \\
& Depth SROCC & 0.927 & 0.924 & 0.892 & 0.964 & 0.819 & 0.885 & 0.902 & 0.936 & 0.906 \\
& Model Size &  94.25 & 61.29 & 163.11 & 187.56 & 101.73 & 81.30 & 87.52 & 91.16 & 108.49 \\
\midrule
{DropGaussian} & SSIM & 0.855 & 0.829 & 0.836 & 0.877 & 0.708 & 0.617 & 0.952 & 0.915 & 0.824 \\
& PSNR & 25.29 & 26.09 & 27.98 & 25.25 & 19.50 & 18.60 & 29.88 & 26.50 & 24.89 \\
& LPIPS & 0.137 & 0.164 & 0.144 & 0.134 & 0.182 & 0.226 & 0.083 & 0.112 & 0.148 \\
& Video SSIM & 0.900 & 0.912 & 0.906 & 0.908 & 0.815 & 0.726 & 0.955 & 0.951 & 0.884 \\
& Video PSNR & 27.15 & 28.94 & 30.82 & 25.13 & 21.60 & 20.41 & 29.08 & 28.56 & 26.46 \\
& Video LPIPS & 0.091 & 0.089 & 0.095 & 0.103 & 0.141 & 0.179 & 0.074 & 0.050 & 0.103 \\
& Depth SROCC & 0.920 & 0.923 & 0.904 & 0.967 & 0.831 & 0.857 & 0.895 & 0.943 & 0.905 \\
& Model Size &  87.36 & 63.73 & 158.90 & 185.88 & 104.39 & 86.98 & 87.75 & 83.79 & 107.35 \\

\midrule
{AD-GS} & SSIM & 0.856 & 0.836 & 0.853 & 0.878 & 0.712 & 0.633 & 0.957 & 0.917 & 0.830 \\
& PSNR & 25.12 & 25.96 & 28.51 & 25.17 & 19.48 & 18.74 & 30.53 & 26.64 & 25.02 \\
& LPIPS & 0.143 & 0.139 & 0.147 & 0.123 & 0.170 & 0.221 & 0.075 & 0.118 & 0.142 \\
& Video SSIM & 0.903 & 0.918 & 0.920 & 0.903 & 0.820 & 0.747 & 0.963 & 0.955 & 0.891 \\
& Video PSNR & 27.39 & 29.09 & 31.53 & 24.72 & 21.69 & 20.78 & 30.07 & 28.80 & 26.76 \\
& Video LPIPS & 0.096 & 0.075 & 0.093 & 0.106 & 0.131 & 0.185 & 0.063 & 0.050 & 0.100 \\
& Depth SROCC & 0.940 & 0.938 & 0.910 & 0.970 & 0.804 & 0.909 & 0.907 & 0.949 & 0.916 \\
& Model Size &  82.36 & 69.97 & 94.30 & 119.47 & 157.01 & 93.75 & 38.78 & 64.22 & 89.98 \\

\bottomrule
\end{tabular}
\end{table*}

\begin{table*}[ht]
\centering
\caption{Scene-wise Quantitative Results on Mip-NeRF360 dataset with 12 input views.}
\vspace{0.5em}
\begin{tabular}{llccccccccc}
\toprule
\textbf{Method} & \textbf{Metric} & \textbf{bicycle} & \textbf{bonsai} & \textbf{counter} & \textbf{garden} & \textbf{kitchen} & \textbf{room} & \textbf{stump} & \textbf{Average} \\
\midrule

{3DGS} 
& SSIM & 0.274 & 0.570 & 0.565 & 0.484 & 0.696 & 0.683 & 0.200 & 0.496 \\
& PSNR & 15.84 & 16.95 & 17.13 & 18.29 & 18.95 & 19.17 & 15.19 & 17.36 \\
& LPIPS & 0.511 & 0.383 & 0.372 & 0.361 & 0.290 & 0.321 & 0.582 & 0.403 \\
& Video SSIM & 0.315 & 0.575 & 0.613 & 0.556 & 0.755 & 0.666 & 0.219 & 0.528 \\
& Video PSNR & 16.42 & 17.11 & 18.48 & 20.24 & 20.56 & 20.26 & 15.64 & 18.39 \\
& Video LPIPS & 0.493 & 0.378 & 0.325 & 0.327 & 0.227 & 0.336 & 0.565 & 0.379 \\
& Depth SROCC & 0.615 & 0.364 & 0.527 & 0.677 & 0.682 & 0.693 & 0.199 & 0.537 \\
& Model Size   & 552.27 & 117.98 & 135.12 & 608.89 & 255.01 & 183.92 & 429.66 & 326.12 \\

\midrule

{FSGS} 
& SSIM & 0.372 & 0.646 & 0.594 & 0.504 & 0.711 & 0.735 & 0.267 & 0.547 \\
& PSNR & 18.18 & 18.91 & 18.01 & 19.01 & 19.16 & 20.47 & 15.81 & 18.51 \\
& LPIPS & 0.564 & 0.352 & 0.381 & 0.408 & 0.269 & 0.297 & 0.607 & 0.411 \\
& Video SSIM & 0.460 & 0.641 & 0.630 & 0.572 & 0.756 & 0.709 & 0.326 & 0.585 \\
& Video PSNR & 18.83 & 19.02 & 19.09 & 20.46 & 20.42 & 21.07 & 16.31 & 19.31 \\
& Video LPIPS & 0.510 & 0.336 & 0.331 & 0.364 & 0.223 & 0.306 & 0.580 & 0.379 \\
& Depth SROCC & 0.885 & 0.801 & 0.743 & 0.924 & 0.917 & 0.822 & 0.424 & 0.788 \\
& Model Size   & 66.33 & 37.10 & 34.75 & 264.53 & 169.08 & 64.03 & 74.62 & 101.49 \\

\midrule

{CoR-GS} 
& SSIM & 0.393 & 0.673 & 0.635 & 0.524 & 0.722 & 0.771 & 0.333 & 0.579 \\
& PSNR & 18.84 & 19.94 & 19.08 & 19.64 & 19.23 & 21.43 & 17.77 & 19.42 \\
& LPIPS & 0.572 & 0.349 & 0.361 & 0.420 & 0.269 & 0.277 & 0.621 & 0.410 \\
& Video SSIM & 0.492 & 0.672 & 0.677 & 0.602 & 0.793 & 0.765 & 0.417 & 0.631 \\
& Video PSNR & 19.81 & 19.79 & 19.99 & 21.71 & 21.23 & 21.98 & 18.82 & 20.48 \\
& Video LPIPS & 0.512 & 0.327 & 0.310 & 0.369 & 0.202 & 0.272 & 0.588 & 0.369 \\
& Depth SROCC & 0.596 & 0.727 & 0.838 & 0.771 & 0.890 & 0.726 & 0.299 & 0.692 \\
& Model Size &70.11 & 32.52 & 38.25 & 283.76 & 169.32 & 72.07 & 46.82 & 101.84 \\

\midrule

{DropGaussian} 
& SSIM & 0.397 & 0.629 & 0.629 & 0.534 & 0.742 & 0.749 & 0.342 & 0.575 \\
& PSNR & 19.08 & 18.01 & 18.45 & 20.07 & 19.45 & 20.67 & 18.51 & 19.18 \\
& LPIPS & 0.563 & 0.363 & 0.368 & 0.412 & 0.259 & 0.300 & 0.616 & 0.412 \\
& Video SSIM & 0.498 & 0.637 & 0.671 & 0.608 & 0.790 & 0.750 & 0.420 & 0.625 \\
& Video PSNR & 20.05 & 18.46 & 19.27 & 21.56 & 20.88 & 21.75 & 19.39 & 20.19 \\
& Video LPIPS & 0.504 & 0.335 & 0.313 & 0.353 & 0.208 & 0.278 & 0.576 & 0.367 \\
& Depth SROCC & 0.664 & 0.480 & 0.638 & 0.886 & 0.778 & 0.698 & 0.341 & 0.641 \\
& Model Size & 76.06 & 34.99 & 33.46 & 277.30 & 170.65 & 62.41 & 53.01 & 101.13 \\

\midrule

{AD-GS (10k)} 
& SSIM & 0.397 & 0.691 & 0.667 & 0.562 & 0.746 & 0.770 & 0.318 & 0.593 \\
& PSNR & 19.04 & 19.99 & 19.51 & 20.36 & 19.78 & 21.42 & 17.52 & 19.66 \\
& LPIPS & 0.550 & 0.315 & 0.334 & 0.381 & 0.251 & 0.278 & 0.595 & 0.386 \\
& Video SSIM & 0.495 & 0.695 & 0.716 & 0.643 & 0.804 & 0.775 & 0.391 & 0.646 \\
& Video PSNR & 19.93 & 20.05 & 20.81 & 22.34 & 21.41 & 22.37 & 18.32 & 20.75 \\
& Video LPIPS & 0.483 & 0.291 & 0.274 & 0.324 & 0.192 & 0.263 & 0.570 & 0.342 \\
& Depth SROCC & 0.881 & 0.762 & 0.858 & 0.885 & 0.920 & 0.764 & 0.489 & 0.794 \\
& Model Size  &154.71 & 49.92 & 38.91 & 223.70 & 101.94 & 59.33 & 138.81 & 109.62 \\

\midrule
{AD-GS (30k)} 
& SSIM & 0.386 & 0.684 & 0.659 & 0.547 & 0.734 & 0.762 & 0.299 & 0.581 \\
& PSNR & 18.505 & 19.449 & 19.319 & 20.047 & 19.765 & 21.203 & 16.957 & 19.321 \\
& LPIPS & 0.531 & 0.313 & 0.333 & 0.382 & 0.254 & 0.281 & 0.592 & 0.384 \\
& Video SSIM & 0.478 & 0.687 & 0.708 & 0.630 & 0.787 & 0.767 & 0.367 & 0.632 \\
& Video PSNR & 19.386 & 19.630 & 20.615 & 21.964 & 21.279 & 22.168 & 17.639 & 20.383 \\
& Video LPIPS & 0.473 & 0.295 & 0.275 & 0.325 & 0.203 & 0.269 & 0.569 & 0.344 \\
& Depth SROCC & 0.893 & 0.797 & 0.855 & 0.842 & 0.931 & 0.803 & 0.522 & 0.806 \\
& Model Size & 178.379 & 45.849 & 37.362 & 214.114 & 90.060 & 52.885 & 163.590 & 111.748 \\

\bottomrule
\end{tabular}
\end{table*}

\begin{table*}[ht]
\centering
\caption{Scene-wise Quantitative Results on Mip-NeRF360 dataset with 24 input views.}
\vspace{0.5em}
\begin{tabular}{llcccccccccc}
\toprule
\textbf{Method} & \textbf{Metric} & \textbf{} & \textbf{bicycle} & \textbf{bonsai} & \textbf{counter} & \textbf{garden} & \textbf{kitchen} & \textbf{room} & \textbf{stump} & \textbf{Average} \\
\midrule
{3DGS} & SSIM &  & 0.437 & 0.792 & 0.750 & 0.731 & 0.826 & 0.817 & 0.550 & 0.701 \\
& PSNR &  & 19.38 & 21.98 & 21.96 & 23.00 & 22.57 & 23.98 & 21.56 & 22.06 \\
& LPIPS &  & 0.405 & 0.213 & 0.237 & 0.184 & 0.174 & 0.207 & 0.349 & 0.253 \\
& Video SSIM &  & 0.515 & 0.849 & 0.778 & 0.776 & 0.876 & 0.810 & 0.583 & 0.741 \\
& Video PSNR &  & 20.68 & 25.46 & 23.35 & 25.17 & 24.58 & 26.05 & 21.70 & 23.85 \\
& Video LPIPS &  & 0.370 & 0.161 & 0.203 & 0.190 & 0.127 & 0.215 & 0.356 & 0.232 \\
& Depth SROCC &  & 0.907 & 0.749 & 0.833 & 0.906 & 0.838 & 0.788 & 0.582 & 0.800 \\
& Model Size &  & 835.78 & 240.80 & 281.27 & 1276.34 & 619.31 & 363.06 & 550.46 & 595.29 \\
\midrule
{FSGS} & SSIM &  & 0.472 & 0.825 & 0.779 & 0.725 & 0.842 & 0.843 & 0.554 & 0.720 \\
& PSNR &  & 20.85 & 24.38 & 23.23 & 23.57 & 23.19 & 24.68 & 21.91 & 23.11 \\
& LPIPS &  & 0.493 & 0.212 & 0.222 & 0.223 & 0.158 & 0.191 & 0.426 & 0.275 \\
& Video SSIM &  & 0.587 & 0.871 & 0.796 & 0.772 & 0.884 & 0.837 & 0.637 & 0.769 \\
& Video PSNR &  & 22.46 & 27.33 & 24.18 & 25.47 & 24.99 & 27.06 & 23.16 & 24.95 \\
& Video LPIPS &  & 0.423 & 0.152 & 0.189 & 0.208 & 0.117 & 0.191 & 0.387 & 0.238 \\
& Depth SROCC &  & 0.918 & 0.904 & 0.924 & 0.964 & 0.926 & 0.865 & 0.752 & 0.893 \\
& Model Size &  & 144.09 & 153.78 & 173.73 & 774.98 & 574.20 & 221.42 & 173.55 & 316.54 \\

\midrule
{CoR-GS} & SSIM &  & 0.478 & 0.831 & 0.783 & 0.733 & 0.856 & 0.855 & 0.560 & 0.728 \\
& PSNR &  & 20.46 & 24.39 & 23.01 & 23.88 & 23.51 & 25.18 & 22.00 & 23.20 \\
& LPIPS &  & 0.501 & 0.215 & 0.220 & 0.222 & 0.147 & 0.180 & 0.422 & 0.272 \\
& Video SSIM &  & 0.596 & 0.884 & 0.818 & 0.787 & 0.901 & 0.877 & 0.656 & 0.788 \\
& Video PSNR &  & 22.04 & 28.04 & 24.45 & 25.93 & 25.34 & 28.04 & 23.78 & 25.37 \\
& Video LPIPS &  & 0.422 & 0.149 & 0.175 & 0.204 & 0.103 & 0.159 & 0.379 & 0.227 \\
& Depth SROCC &  & 0.491 & 0.848 & 0.868 & 0.918 & 0.958 & 0.852 & 0.760 & 0.814 \\
& Model Size &  & 150.84 & 151.81 & 183.19 & 848.38 & 550.98 & 242.64 & 198.85 & 332.38 \\

\midrule
{DropGaussian} & SSIM &  & 0.487 & 0.821 & 0.789 & 0.737 & 0.861 & 0.851 & 0.577 & 0.732 \\
& PSNR &  & 21.46 & 24.24 & 23.12 & 23.75 & 23.52 & 24.79 & 22.10 & 23.28 \\
& LPIPS &  & 0.494 & 0.227 & 0.227 & 0.233 & 0.147 & 0.187 & 0.425 & 0.277 \\
& Video SSIM &  & 0.607 & 0.871 & 0.822 & 0.803 & 0.909 & 0.870 & 0.673 & 0.794 \\
& Video PSNR &  & 23.22 & 27.38 & 24.53 & 26.41 & 25.73 & 27.56 & 24.03 & 25.55 \\
& Video LPIPS &  & 0.419 & 0.161 & 0.173 & 0.193 & 0.096 & 0.159 & 0.372 & 0.225 \\
& Depth SROCC &  & 0.709 & 0.846 & 0.862 & 0.946 & 0.838 & 0.806 & 0.770 & 0.825 \\
& Model Size &  & 140.76 & 132.16 & 161.01 & 861.39 & 572.59 & 223.52 & 190.31 & 325.96 \\

\midrule
{AD-GS (10k)} & SSIM &  & 0.500 & 0.852 & 0.805 & 0.768 & 0.863 & 0.855 & 0.605 & 0.750 \\
& PSNR &  & 21.16 & 25.11 & 23.56 & 24.29 & 23.63 & 25.19 & 22.79 & 23.68 \\
& LPIPS &  & 0.462 & 0.188 & 0.201 & 0.192 & 0.143 & 0.183 & 0.375 & 0.249 \\
& Video SSIM &  & 0.625 & 0.902 & 0.841 & 0.831 & 0.911 & 0.888 & 0.701 & 0.814 \\
& Video PSNR &  & 23.04 & 28.37 & 24.83 & 26.87 & 25.56 & 28.30 & 24.54 & 25.93 \\
& Video LPIPS &  & 0.379 & 0.120 & 0.152 & 0.159 & 0.095 & 0.146 & 0.332 & 0.198 \\
& Depth SROCC &  & 0.926 & 0.916 & 0.930 & 0.966 & 0.959 & 0.866 & 0.779 & 0.906 \\
& Model Size &  & 306.90 & 103.86 & 116.84 & 516.38 & 184.05 & 150.93 & 241.14 & 231.44 \\

\midrule
{AD-GS (30k)} 
& SSIM &  & 0.498 & 0.855 & 0.798 & 0.762 & 0.859 & 0.851 & 0.591 & 0.745 \\
& PSNR &  & 20.843 & 24.869 & 23.457 & 23.906 & 23.509 & 25.142 & 22.307 & 23.433 \\
& LPIPS &  & 0.435 & 0.176 & 0.198 & 0.186 & 0.143 & 0.186 & 0.367 & 0.242 \\
& Video SSIM &  & 0.617 & 0.909 & 0.830 & 0.824 & 0.906 & 0.880 & 0.683 & 0.807 \\
& Video PSNR &  & 22.462 & 28.408 & 24.581 & 26.598 & 25.478 & 28.202 & 24.128 & 25.694 \\
& Video LPIPS &  & 0.358 & 0.106 & 0.158 & 0.158 & 0.098 & 0.157 & 0.334 & 0.196 \\
& Depth SROCC &  & 0.928 & 0.917 & 0.930 & 0.967 & 0.956 & 0.880 & 0.769 & 0.907 \\
& Model Size &  & 325.77 & 87.28 & 95.71 & 434.75 & 150.49 & 113.78 & 234.51 & 206.04 \\

\bottomrule
\end{tabular}
\end{table*}

\clearpage
\newpage
\twocolumn


\bibliographystyle{ACM-Reference-Format}
\bibliography{software}

\end{document}